\documentclass[sn-mathphys-num]{sn-jnl}

\usepackage{graphicx}%
\usepackage{multirow}%
\usepackage{amsmath,amssymb,amsfonts}%
\usepackage{amsthm}%
\usepackage{mathrsfs}%
\usepackage[title]{appendix}%
\usepackage{xcolor}%
\usepackage{textcomp}%
\usepackage{manyfoot}%
\usepackage{booktabs}%
\usepackage{algorithm}%
\usepackage{algorithmicx}%
\usepackage{algpseudocode}%
\usepackage{listings}%
\usepackage{tabularx}
\usepackage{dcolumn}

\usepackage{pythonhighlight}
\usepackage{enumitem}
\usepackage{tcolorbox}
\usepackage{booktabs}

\newcolumntype{Y}{>{\raggedright\arraybackslash}X} 
\newcolumntype{R}{>{\raggedleft\arraybackslash}X}  

\usepackage{subfig}
\usepackage{graphicx}

\theoremstyle{thmstyleone}%
\theoremstyle{thmstyletwo}%

\theoremstyle{thmstylethree}%

\begin{document}

\title[A case study on the transformative potential of AI in software engineering]{A case study on the transformative potential of AI in software engineering on LeetCode and ChatGPT}


\author[1]{\fnm{Manuel} \sur{Merkel}}\email{st155131@stud.uni-stuttgart.de}

\author*[2,3,4]{\fnm{Jens} \sur{D\"orpinghaus}}\email{doerpinghaus@uni-koblenz.de}

\affil[1]{\orgname{University of Stuttgart}, \city{Stuttgart},  \country{Germany}}

\affil[2]{\orgname{Federal Institute for Vocational Education and Training (BIBB)}, \city{Bonn}, \country{Germany}}

\affil[3]{\orgname{University of Koblenz}, \city{Koblenz}, \country{Germany}}

\affil[4]{\orgdiv{Department of Computer Science and Media Technology}, \orgname{Linnaeus University}, \city{V\"axj\"o},  \country{Sweden}}


\abstract{The recent surge in the field of generative artificial intelligence (GenAI) has the potential to bring about transformative changes across a range of sectors, including software engineering and education. As GenAI tools, such as OpenAI's ChatGPT, are increasingly utilised in software engineering, it becomes imperative to understand the impact of these technologies on the software product. This study employs a methodological approach, comprising web scraping and data mining from LeetCode, with the objective of comparing the software quality of Python programs produced by LeetCode users with that generated by GPT-4o. In order to gain insight into these matters, this study addresses the question whether GPT-4o produces software of superior quality to that produced by humans.


The findings indicate that GPT-4o does not present a considerable impediment to code quality, understandability, or runtime when generating code on a limited scale. Indeed, the generated code even exhibits significantly lower values across all three metrics in comparison to the user-written code. However, no significantly superior values were observed for the generated code in terms of memory usage in comparison to the user code, which contravened the expectations. Furthermore, it will be demonstrated that GPT-4o encountered challenges in generalising to problems that were not included in the training data set. 

This contribution presents a first large-scale study comparing generated code with human-written code based on LeetCode platform based on multiple measures including code quality, code understandability, time behaviour and resource utilisation. All data is publicly available for further research.}

\keywords{keyword1, Keyword2, Keyword3, Keyword4}



\maketitle

\section{Introduction}\label{sec1}

\subsection{Background}

The question of whether machines can think was first posed as early as 1950, making it almost 75 years old. This illustrates the historical dimension of the topic of artificial intelligence. However, it was only with the release of ChatGPT as a free chatbot that a real boom in GenAI was triggered in 2022 (time to reach 1 million users of ChatGPT\footnote{Time to reach 1 million ChatGPT users, \url{https://explodingtopics.com/blog/chatgpt-users}}), which, in turn, gave rise to a plethora of GenAI applications in the subsequent period. These GenAI tools generate creative content or information through a conversational approach in a format that humans can understand. The functionality of GenAI tools is based on large language models (LLMs), which have been trained on an enormous amount of data from the Internet~\cite{villalobos2024rundatalimitsllm}. Consequently, these tools have access to a vast repository of information on already published data, which could make them a potentially powerful tool. In addition to its application in creative processes, such as image generation, GenAI is also utilised in computer science, particularly in the field of software development. The potential range of applications is wide-ranging and has the capacity to affect a significant restructuring of the development process~\cite{10176168}. In order to illustrate this point, it is useful to consider the Stack Overflow\footnote{Stack Overflow, \url{https://stackoverflow.com/}} platform, which is aimed at software developers. Over an extended period, it has frequently been employed to address issues pertaining to difficulties in comprehending code, locating errors, and asking general questions. As a result, it has become a crucial component of the software engineering process for developers. Nevertheless, the advent of GenAI tools such as ChatGPT has the potential to significantly reduce the time required for this process. ChatGPT generates an answer based on prompts provided by the developer, thus obviating the need for the developer to collate information from online sources, such as Stack Overflow. Furthermore, ChatGPT is capable of both composing and executing code. This renders it appropriate for the purposes of debugging code, generating code from natural language, elucidating code and responding to questions on general topics~\cite{10176168}. Therefore, it can be employed as a kind of \textit{pair programmer}~\cite{ma2023aibetterprogrammingpartner} or as a kind of \textit{tutor}~\cite{10.1145/3626252.3630938} for all, including both beginners and advanced users. While some have even gone so far as to claim that ``[j]obs such as coders, software developers, computer programmers, and data scientists are at risk of being displaced by AI [...]''~\cite{zarifhonarvar2024economics}, others have expressed the opinion that GenAI tools ``[...] did not end up being the crazy productivity booster that I thought it would be, because programming is designing and these tools aren't good enough (yet) to assist me with this seriously''~\cite{ludic2023}. Indeed, empirical studies have been conducted to investigate the impact of GenAI tools on the software development process. In order to evaluate the efficacy of GenAI, a series of benchmarks~\cite{chen2021evaluatinglargelanguagemodels, lu2021codexgluemachinelearningbenchmark, hendrycks2021measuringcodingchallengecompetence, austin2021programsynthesislargelanguage} were devised to initially assess its basic functional correctness~\cite{Nguyen, Liu}. Subsequent studies have examined software quality in terms of poor code patterns~\cite{Siddiq2022} and code understandability~\cite{LiuSQ}, while others have analysed the security of the code~\cite{9833571} and errors therein~\cite{10174227}. Nevertheless, the studies do not proceed to a subsequent stage, namely a comparison of the quality of the generated software with that of the code written by humans. Consequently, it is unclear whether the use of GenAI has a positive effect on software development and the resulting code quality. Code generation may prove beneficial for developers, but it can also result in the inadvertent introduction of suboptimal code quality in production. This can become a significant challenge as it necessitates the involvement of additional developers to maintain the code~\cite{jones2011economics}. This study aims to explore the basis for comparing quality of Python code generated by GenAI with that of code produced by humans. 

For this, foundations regarding the produced software quality of GenAI tools are inspected. The study aims to identify potential challenges and opportunities associated with the adoption of GenAI, providing insights for educators, policymakers, and developers. In order to achieve the aforementioned objectives, this study addresses the following research question: \textit{``Does GenAI produce better software quality than humans?''}

\subsection{Research Question}
The research question is addressed using a large-scale methodology, which is characterised by a relatively low degree of control. Data science methods such as web scraping~\cite{WebScraping} of coding problems from LeetCode\footnote{LeetCode, \url{https://leetcode.com/}} are used to provide a solid basis for evaluating the impact of GenAI on software quality on a large scale. The LeetCode platform currently provides access to over 2,992 coding problems, offering a substantial repository of information for researchers and developers alike. In addition to these problems, the platform features over 278,397 posts, manually written and published by users. By employing a range of web scraping and data mining techniques~\cite{MSR}, the following information has been extracted:

\begin{itemize}
    \item The static code analysis tool SonarQube\footnote{SonarQube, \url{https://www.sonarsource.com/products/sonarqube/}} is employed to report on two key aspects of software quality: number of code smells per line of code, which determine \textit{code quality}, and cognitive complexity per line of code, which represents \textit{code understandability}.
    \item The LeetCode platform is employed to report on two key aspects of performance efficiency: runtime, which determines \textit{time behaviour}, and memory usage, which represents \textit{resource utilisation}.
\end{itemize}

\subsection{Hypotheses}
The quality of software can be evaluated based on a number of different characteristics. This large-scale research sought to ascertain whether the generated code of GenAI exhibits superior software quality when compared to manually created code, regarding to four quality metrics. In order to achieve this objective, four hypotheses have been formulated regarding the four quality characteristics, which can be seen in Table \ref{tab:HypothesisRQ1}. The motivation behind these hypotheses was based on the assumption that GenAI already has an extensive set of training data, ranging from various algorithms to different rules of software quality. The initial estimates and assumptions posit that the publicly available, high-quality training data from the Internet will soon be depleted~\cite{villalobos2024rundatalimitsllm}, underscoring the vast scale of the data sets. This could provide GenAI with more information about software development than the average individual~\cite{Finnie-AnsleyCS1}. Moreover, the existing literature indicates that the GenAI's capability to correctly resolve coding problems is already noteworthy. A comprehensive overview can be found in Section \ref{sec:relworkWD}. However, the objective was to ascertain whether the optimal software quality can be generated by utilising the GenAI, and whether this quality is superior to that produced by humans. This was evaluated in terms of code quality, code understandability, resource utilisation and execution time.

\begin{table}[h]
    \centering
    \scriptsize
    \begin{tabularx}{\textwidth}{p{2.3cm} X X}
        \toprule
        & \textbf{Null hypothesis} & \textbf{Alternative Hypothesis}\\
        \midrule
        \textbf{Code Quality}
        & H$_{0}^{1}$ GenAI produces \textit{less or equal} code quality than developers on LeetCode.
        & H$_{1}^{1}$ GenAI produces \textit{better} code quality than developers on LeetCode.
        \\\hline
        \textbf{Code Understandability}
        & H$_{0}^{2}$ GenAI produces \textit{less or equal} code understandability than developers on LeetCode.
        & H$_{1}^{2}$ GenAI produces \textit{better} code understandability than developers on LeetCode.
        \\\hline
        \textbf{Resource Utilisation}
        & H$_{0}^{3}$ GenAI produces code that utilises \textit{equal or more} resources than developers on LeetCode.
        & H$_{1}^{3}$ GenAI produces code that utilises \textit{less} resources than developers on LeetCode.
        \\\hline
        \textbf{Time Behaviour}
        & H$_{0}^{4}$ GenAI produces code that takes \textit{equal or more} time to run than developers on LeetCode.
        & H$_{1}^{4}$ GenAI produces code that takes \textit{less} time to run than developers on LeetCode.
        \\
        \bottomrule
    \end{tabularx}
    \caption{Null hypotheses with their alternative hypotheses for the research question}
    \label{tab:HypothesisRQ1}
\end{table}

\subsection{Structure}

This article comprises six sections. The initial section offers an overview, followed by a section covering current research and related work. The third part outlines the methodology employed in the study, its objectives and data mining approaches.  The fourth part presents experimental results and discusses them together with the hypotheses. After a detailed discussion, the conclusion and an outlook are outlined in the final section.

\section{Related Work}

The increasing prevalence and rapid dissemination of GenAI tools have highlighted the necessity for a comprehensive scientific evaluation within the field of computer science. 
The methodology of this study addresses web scraping and data mining of generated code. Consequently, the related work on this topic is discussed in detail. 
This section is followed by a summary, which presents the research gap and elucidates the motivation for the present study.

The topic of computer science education, see for example \cite{kandlhofer2016artificial}, the usage of artificial intelligence in computer science occupations \cite{felten2019occupational,udelhofen2024professionals,tredinnick2017artificial,incollectionderksen,kostadinovska2024educational}, and AI approaches for labor market research \cite{dorpinghaus2023challenges,fischer2024web} and data mining within computational social sciences are only marginally related, see for example \cite{bittermann2024natural,dorpinghaus2023towards,dorpinghaus2020knowledge,Dorpinghaus2022-hk,fechner2023classifying,hein2024linked,reiser2024towards}.

\subsection{Web Scraping and Data Mining}\label{sec:relworkWD}
The practice of web scraping~\cite{WebScraping} and subsequent data mining~\cite{MSR} has already become a well-established methodology for the evaluation and comparison of software quality. The wealth of data and information available on the Internet has a variety of evaluation options. In this context, the following sections presents an overview of the platforms and methods that are currently being used for the evaluation of GenAI. The following Table \ref{tab:RelWorkWebSc} provides an overview of the related work presented in this context. The table includes only those scientific papers that were subjected to a more detailed description. The papers have been categorised in accordance with the respective sections. The respective research areas are presented in the table; the results can then be found in the section where the studies are elucidated.

\begin{table}[h]
    \centering
    \scriptsize
    \begin{tabularx}{\textwidth}{>{\raggedright\arraybackslash}p{2.3cm} >{\raggedright\arraybackslash}p{1.6cm} >{\raggedright\arraybackslash}p{2.3cm} X}
        \toprule
        \textbf{Authors} & \textbf{Platform} & \textbf{GenAI} & \textbf{Research Area} \\
        \midrule
        Ray et al.~\cite{RayMSR} & GitHub & -- & Comparing the software quality of different programming languages (PLs)\\
        Berger et al.~\cite{BergerReplication} & GitHub & -- & Replication study of Ray et al's study\\
        Yu et al.~\cite{GitHubMSRGenerated} & GitHub & GitHub Copilot & Comparison of generated code with human-written code \& assessment of code understandability \\
        \midrule
        Siddiq et al.~\cite{Siddiq2022} & HumanEval & GPT Code Clippy \& GitHub Copilot & Analysis of security \& maintainability code smells in training data and generated code from GenAIs\\
        \midrule
        Nguyen and Nadi~\cite{Nguyen} & LeetCode & GitHub Copilot & Comparison of generated code of 4 PLs for functional correctness and code understandability \\
        Liu et al.~\cite{Liu} & LeetCode \& CWEs & ChatGPT & Comparison of generated code of 5 PLs for functional correctness, code understandability \& code security (with multi-round fixing)\\
        Liu et al.~\cite{LiuSQ} & LeetCode & ChatGPT & Comparison of generated code of 2 PLs for functional correctness, maintainability \& reliability (with multi-round fixing)\\
        Idrisov and Schlippe~\cite{Idrisov} & LeetCode & 6 GenAIs & Comparison of generated code of 6 GenAIs for functional correctness\\
        Liu et al.~\cite{LiuGPTs} & HumanEval+ & 26 GenAIs & Comparison of generated code of 26 GenAIs for functional correctness\\
        Coignion et al.~\cite{Coignion} & LeetCode & 18 GenAIs & Comparison of generated code of 18 GenAIs (also with user solutions) for performance efficiency \\
        \bottomrule
    \end{tabularx}
    \caption{Overview of related work on web scraping and data mining discussed in detail}
    \label{tab:RelWorkWebSc}
\end{table}

\subsection{Data Mining on GitHub}
The GitHub\footnote{GitHub, \url{https://github.com/home}} platform has frequently been employed as a foundation for data collection, given its vast archive of publicly accessible information spanning an extended period. To illustrate, Ray et al.~\cite{RayMSR} conducted an analysis of 729 projects in 2014, gathering and examining data on the project history, project size, and team size across 17 distinct programming languages. The objective of the study was to examine the impact of programming languages on software quality. The findings indicate that language design exerts a minor influence on software quality. However, when Berger et al.~\cite{BergerReplication} conducted a replication study in 2019, they were unable to achieve the desired outcome with respect to two of the four research questions. Among other issues, the data set was not entirely suitable for some of the analysis, including the selection of TypeScript projects created prior to 2012 (i.e., before its official release). Furthermore, the practical effect size for the results of Ray et al. is so small that the study's conclusion could not be adequately supported. Nevertheless, the study by Ray et al.~\cite{RayMSR} has made an important contribution to the field of mining software repositories. The methods were adopted in many cases~\cite{GaoMSR, KocharMSR} and modified~\cite{BognerMerkelMSR} in accordance with the conclusions of preceding studies. The utilisation of GitHub as a rich data source for the evaluation of the quality of generated code presents a significant challenge at the present time. This is attributable to the nascent state of research in the domain of GenAI, coupled with the inherent complexity of recognising the generated code to be analysed~\cite{AICodeDetector}.

In 2024, however, Yu et al.~\cite{GitHubMSRGenerated} developed a method to access the generated code via GitHub. By searching for specific keywords within the comments of the code base, such as ``generated by ChatGPT'', the authors were able to track and analyse the associated generated code. In addition to the general information on the characteristics of the projects that contained generated code and on the characteristics of the generated code itself, the history of the generated code was also examined and analysed using, among others, the methods provided by Ray et al.~\cite{RayMSR}. Furthermore, the generated code was subjected to comparison with code written by humans. The generated code and the human-written code were examined using SonarQube, a static code analysis tool, to ascertain metrics such as lines of code, cyclomatic complexity, and cognitive complexity, which collectively provide insight into the complexity of the code. The analysis was conducted on 983 projects containing generated code. The authors conclude that developers tend to favour ChatGPT and GitHub Copilot\footnote{GitHub Copilot, \url{https://github.com/features/copilot}} as tools for generating Python and JavaScript code in the context of data processing and transformation. The generated code is generally shorter, is less likely to be subsequently adapted, and therefore contains fewer errors than code written by humans. Finally, for Python and JavaScript, the generated code exhibits a lower median cyclomatic complexity and cognitive complexity than the code written by humans. Interestingly, the opposite is true for C/C++, Java and TypeScript. Further analyses of this insight are planned for future work. The authors were thus among the first and only to analyse the quality of the code generated by LLMs in a large-scale study using GitHub.

\subsection{Data Mining on Benchmarks}\label{sec:benchmarks}
The widely-used GitHub platform has been demonstrated to be of limited utility for the web scraping of generated code. Consequently, an endeavour has been made to devise a series of benchmarks for the analysis, evaluation and comparison of the code output of LLMs. The benchmarks, which comprise a data set of coding problems, include HumanEval~\cite{chen2021evaluatinglargelanguagemodels}, CodeXGLUE~\cite{lu2021codexgluemachinelearningbenchmark}, APPS~\cite{hendrycks2021measuringcodingchallengecompetence} and MBPP~\cite{austin2021programsynthesislargelanguage}, all of which were created in 2021. The available datasets typically comprise a substantial number of problems. The HumanEval benchmark, for example, was developed by OpenAI\footnote{OpenAI, \url{https://openai.com/}} for the purpose of evaluating their Codex model. However, the extensive number of problems of those benchmarks present a problematic aspect: not all of them have been created manually. A portion of the datasets is sourced from GitHub and is highly probable to be included in the AIs' training data~\cite{chen2021evaluatinglargelanguagemodels}. Furthermore, it is important to note, however, that these benchmark data sets are not subject to maintenance and therefore do not receive updates or extensions on a regular basis.

As one of the inaugural studies in the domain of software quality as it pertains to GenAI, the authors Siddiq et al.~\cite{Siddiq2022} undertook in 2022 an in-depth analysis of the code quality of GPT-Code-Clippy\footnote{GPT Code Clippy, \url{https://github.com/CodedotAl/gpt-code-clippy/wiki}}, the open-source version of GitHub Copilot. As part of the investigation, an analysis was conducted on both the Python code samples from the training datasets and the generated code produced by the AI. The generated code was evaluated using a data set of 164 programming problems, which were collected from the HumanEval repository. The training data samples and the generated code for the problems were examined using the static code analysis tools Pylint\footnote{Pylint, \url{https://pylint.readthedocs.io/en/stable/}} and Bandit\footnote{Bandit, \url{https://bandit.readthedocs.io/en/latest/}}. During this process, code smells were documented and classified into two categories: security smells, which indicate potential vulnerabilities in the code, and non-security-related smells, which indicate poor maintainability. The results of the investigation demonstrate that both security-related and maintainability-related smells are present in the training set of the GenAI and are therefore also reflected in the generated output. The latter finding is consistent across both the open-source variant, GPT-Code-Clippy, and the closed-source variant, GitHub Copilot.

\subsection{Data Mining on LeetCode}
However, due to the paucity of updates or the limited size of the data sets in the aforementioned benchmarks, the majority of studies currently employ the extensive programming problems from LeetCode, with new coding problems being added on a regular basis.

In contrast with the study conducted by Siddiq et al.~\cite{Siddiq2022}, Nguyen and Nadi~\cite{Nguyen} examined in 2022 the correctness and understandability of the solutions generated by GitHub Copilot. Moreover, the potential for generating code in multiple programming languages was employed to facilitate comparisons between the various languages. The LeetCode platform was utilised to randomly collect 33 programming problems (four categorised as \textit{easy}, 17 as \textit{medium} and 12 as \textit{hard}). For the selected programming problems, a total of 132 solutions were generated by Copilot in the languages Python, Java, JavaScript, and C. Subsequently, the solutions were submitted to LeetCode in order to evaluate the number of test cases that each solution passes. Furthermore, the code files were analysed using the static code analysis tool SonarQube, which permits a judgement to be made regarding the cognitive complexity of the code and thus an evaluation of its understandability. In the context of \textit{easy} programming problems, Copilot exhibited 100\% correctness across all four languages. However, Java demonstrated the highest overall performance, with a correctness score of 57\%, while JavaScript exhibited the lowest score of 27\%. No significant differences were observed between the languages in terms of code understandability.

In 2024, Liu et al.~\cite{Liu} addressed the same topics regarding generated code as those addressed by Nguyen and Nadi~\cite{Nguyen} in their study - specifically, code correctness and complexity - and additionally considered code security. Furthermore, the study is considerably more extensive, and the utilisation of revised methodologies facilitated a more comprehensive understanding of the subject matter. A total of 728 algorithmic problems were randomly selected from LeetCode, and 18 CWEs\footnote{Common Weakness Enumeration, \url{https://cwe.mitre.org/}} comprising 54 code scenarios were considered. The solutions to the programming problems were generated with ChatGPT (GPT-3.5) for five programming languages: C, C++, Java, Python and JavaScript. The data was analysed according to a temporal framework, with the problems divided into two categories: those published prior to the conclusion of the training data for the GenAI and those published subsequent to this point. As anticipated, ChatGPT demonstrated superior functionality in solving problems presumed to be included in the training data (68.41\%) compared to those not included in the training data (20.27\%). As LeetCode provides feedback on the specific error case, the authors employed a multi-round fixing method. This entailed sending another prompt to ChatGPT up to five times, including the faulty code information, if the code was not functionally correct. The application of this multi-round fixing method resulted in the improvement and functional correctness of 31\% of the incorrectly submitted code solutions. In terms of code understandability, the results differed across the various programming languages, leading to a divergence from the findings of Nguyen and Nadi~\cite{Nguyen} in their study with GitHub Copilot. Liu et al. ascribe this to language-specific properties. Multi-round fixing resulted in either no change or a decline in the level of code understandability. Furthermore, the authors conclude that the generated code of ChatGPT exhibits vulnerabilities, but it is promising in eliminating these vulnerabilities when multi-round fixing the issue (89\% removed).

The study by Liu et al.~\cite{LiuSQ} presents a comprehensive investigation into the performance and quality of code generated by ChatGPT (GPT-3.5) in 2024. The analysis encompasses 4,066 code snippets, generated for 2,033 programming tasks from LeetCode, written in both Java and Python. Accordingly, the study seeks to evaluate the correctness of the generated code, identify common code quality issues through static code analysis, and investigate methods to mitigate these issues through ChatGPT's self-repairing capabilities, in a manner analogous to that described by Liu et al.~\cite{Liu} before. The findings indicate that while ChatGPT produces a considerable proportion of functionally correct code (66\% for Python and 69\% for Java), there is nevertheless a considerable amount of code that suffers from issues such as maintainability problems, erroneous outputs, and runtime errors. The study indicates that ChatGPT's performance deteriorates with the increasing complexity of tasks, exhibiting better results for those categorised as \textit{easy} and declining for those categorised as \textit{medium} and \textit{hard}. Furthermore, while ChatGPT displays some capability for self-repair (through prompting information from static code analysis tools or error information), it frequently introduces new issues during this process. This study corroborates the findings of previous studies by~\cite{Liu} and~\cite{Nguyen}, while also providing new insights into the relationship between code quality and software errors in the generated code by ChatGPT.

Nevertheless, there are many other studies that address the functional correctness of the solutions generated by LeetCode, e.g~\cite{Bucaioni} or~\cite{Doderlein} that also consider the settings of the GenAI model. Notwithstanding, studies such as~\cite{Idrisov} and~\cite{LiuGPTs} have concentrated on a comparative analysis of code solutions generated with disparate AI models in 2024. Idrisov and Schlippe~\cite{Idrisov} utilised the LeetCode problem set, albeit on a smaller scale of 18 problems constructed subsequent to the training dataset. In conclusion, the results demonstrated that GitHub Copilot was the most effective in solving problems, with a success rate of 50\%. BingAI Chat\footnote{Microsoft Copilot, \url{https://www.microsoft.com/en-us/microsoft-copilot/learn?ep=0&form=MA13LV&es=31}} (with GPT-4.0) exhibited a 38.9\% success rate, while ChatGPT (with GPT-3.5) and Meta's Code Llama (with Llama 2) demonstrated a 22.2\% success rate, while StarCoder\footnote{Hugging Face StarCoder, \url{https://huggingface.co/blog/starcoder}} and InstructCodeT5+\footnote{Salesforce InstructCodeT5+, \url{https://huggingface.co/Salesforce/instructcodet5p-16b}} exhibited a 5.6\% success rate. Amazon's CodeWhisperer\footnote{Amazon CodeWhisperer, \url{https://aws.amazon.com/q/developer/?nc1=h_ls}} was unable to solve any of the problems. In contrast, the authors from~\cite{LiuGPTs} employed the HumanEval dataset, expanded it, and conducted an analysis with 26 different GenAIs on it. However, it is possible that the problems may also be included in the training datasets of the various GenAIs. The results indicated that GPT-4 was able to solve the most problems with 76.2\%. In addition to the two different datasets of varying sizes, this result, when considered alongside the results of~\cite{Idrisov}, suggests that GPT-4 may also have difficulties in generalising to unseen problems. A comparison of the two results reveals that Idrisov utilises problems that were created subsequent to the cutoff date of the GPT-4 training dataset (functional correctness of 38.9\%), whereas Liu et al. employ problems that may be included in the training dataset of GPT-4 (functional correctness of 76.2\%). This is a point that has previously been highlighted by Liu et al.~\cite{Liu} in the case of ChatGPT (GPT-3.5).

In their study, Coignion et al.~\cite{Coignion} also conducted a comparative analysis of various LLMs. The principal aim of the study was to evaluate the efficiency of code generation of GenAIs and to compare their performance with that of solutions created by humans. In order to achieve this, programming problems from LeetCode were employed, with particular attention paid to ensuring that none of these were included in the AI training data set in an effort to prevent contamination of the data. The problem set comprised a total of 204 problems, with ten solutions generated in Python for each of the 18 code LLMs. The metric \textit{Memory Usage} and \textit{Runtime}, provided by LeetCode for each uploaded solution, was used as a measure of efficiency. In order to facilitate a comparison of performance, the runtime was measured locally with pytest-benchmark\footnote{\textit{pytest-benchmark}, \url{https://pypi.org/project/pytest-benchmark/}} and compared with this provided by LeetCode. A comparison with solutions created by humans was enabled by determining the respective rank of the generated solutions in terms of runtime, as determined by LeetCode. Overall, the authors concluded that LLMs generate code with comparable performance. Furthermore, the authors conclude that the code of LLMs is, on average, more efficient than code written by humans. However, it should be noted that Bucaioni et al.~\cite{Bucaioni} found the opposite to be true for the generation of Java and C++ solutions to 240 programming problems of LeetCode with ChatGPT (GPT-3.5).

\subsection{Summary}\label{RelWorkSumm}
A review of the literature reveals a considerable number of studies that have analysed and compared the various qualities of LLMs. This encompasses studies that compare different LLMs with regard to their code generation capabilities and studies that contrast the generation of code in different programming languages. In this regard, the LeetCode platform represents a valuable resource for coding problems, offering a consistent stream of new coding problems, a substantial number of test cases, and comprehensive problem descriptions. This makes it an optimal choice for evaluating the capabilities of GenAIs.

However, Yu et al. is one of the few studies to undertake a comparison of generated code with human-written code on a large scale, utilising code from GitHub. Given that some of the GenAIs have undergone training on code sourced from GitHub, there is a possibility of data contamination in their study. Additionally, their data set for Python is considerably smaller than that used in this study, and the authors only analysed code understandability, leaving a gap in further insights into software quality. Furthermore, a direct comparison of the generated code with human-written code for the same problem set was not possible for them. It is the intention to address these shortcomings more comprehensively, to incorporate additional quality attributes into the work, and to facilitate a direct comparison between the generated and the human-written code. Furthermore, a benchmark has been established on which further evaluations can be conducted, thus providing the opportunity to compare the results.

With regard to the performance efficiency of the generated code, a small number of comparisons have been conducted with the code written by humans. However, this study provides valuable insights into the GPT-4o model in a large-scale setting, which has yet to be incorporated into the existing literature. Additionally, some of these studies present conflicting findingss, e.g., Bucaioni et al. and Coignion et al., which require further investigation.

\section{Methodology}\label{sec:webscrap}
\begin{figure}[t]
    \centering
    \includegraphics[width=\columnwidth]{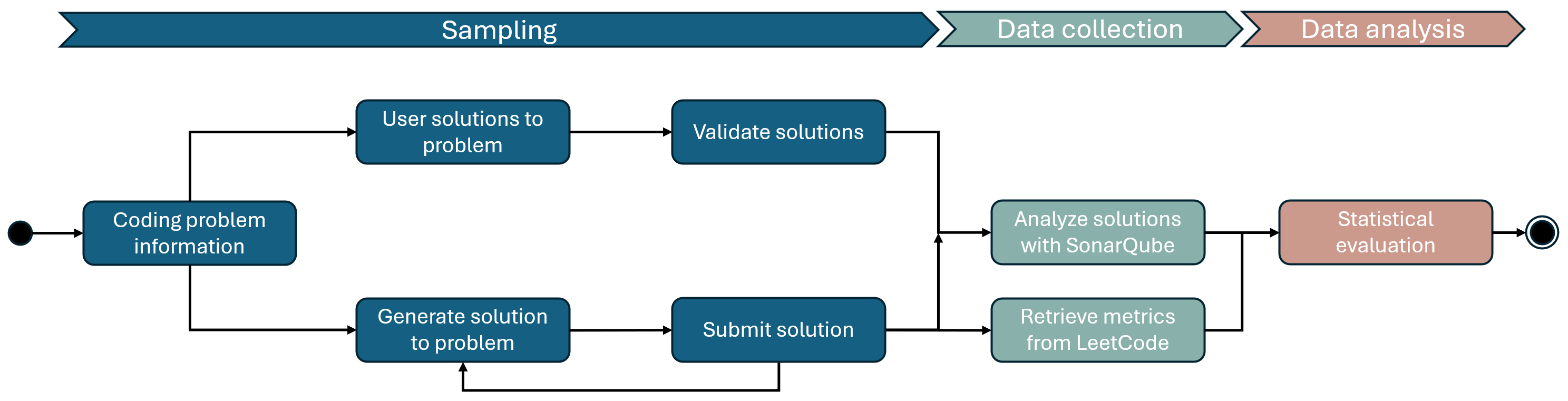}
    \caption{Research process of web scraping and data mining}
    \label{fig:research-process}
\end{figure}

To answer the research question, the method of web scraping was selected as the most appropriate means of gathering the requisite information. The subsequent sections will provide a detailed account of the study objects and the method of data collection employed. Figure \ref{fig:research-process} provides a visual representation of the abstract sequence of the web scraping and data mining process, presented as a flowchart. The initial stage was sampling, which is discussed in greater detail in Section \ref{sec:studobj}. The subsequent Section \ref{sec:datamining} addresses the process of data collection, including the selection of metrics and the methods employed to measure them. The final step is the data analysis, which entails the statistical evaluation of the hypotheses.

\subsection{Study Objects}\label{sec:studobj}
At the outset of the process, a platform for the extraction of information was established with the objective of obtaining the study objects and thus the samples. Subsequently, samples were collected based on the independent variable, which in this case was the code snippets written by humans on the one hand and the generated code snippets by the GenAI on the other. The following sections elucidates this process in greater detail.

\subsubsection{Selection of Platform}

\begin{figure}[t]
    \centering
    \includegraphics[width=\columnwidth]{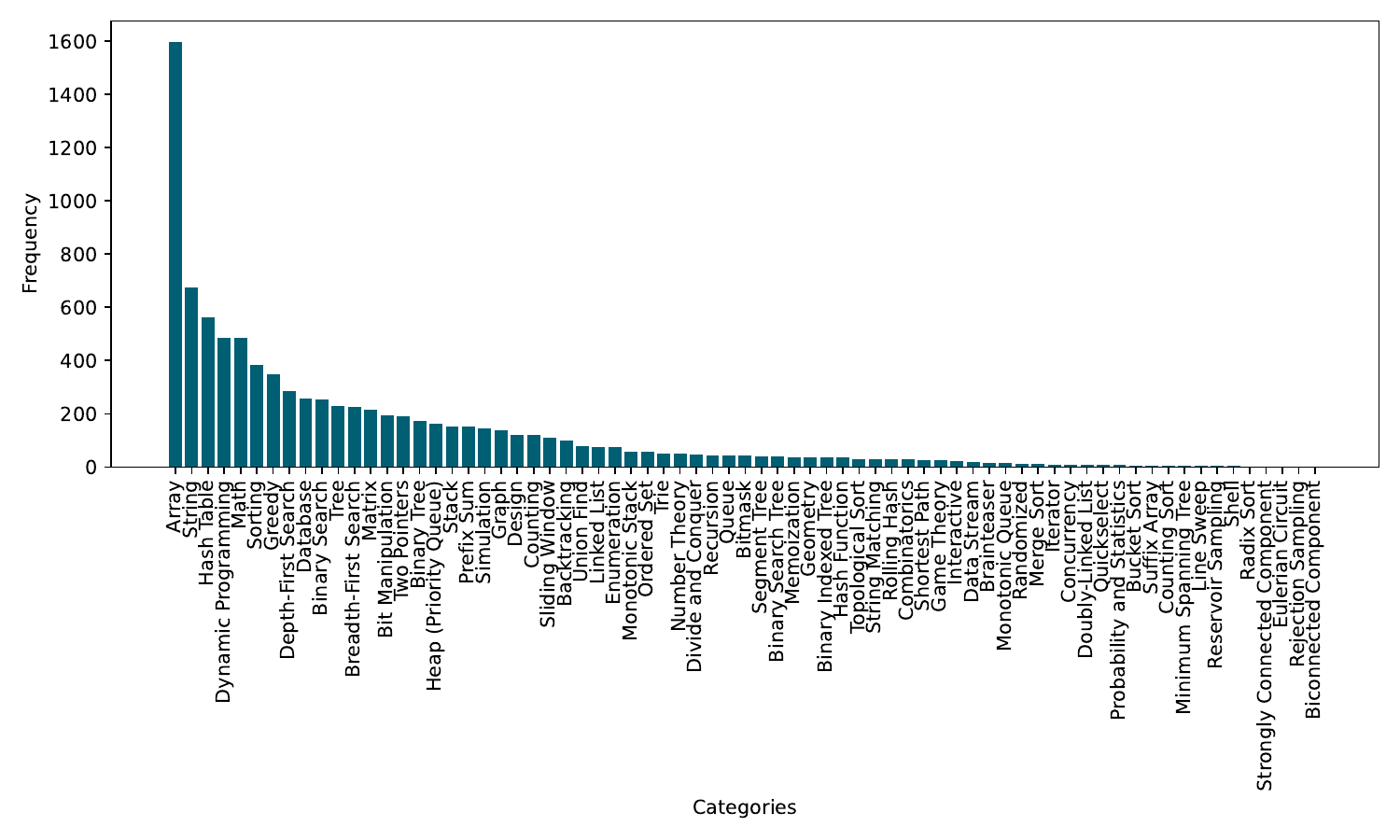}
    \caption{Distribution of problems across categories}
    \label{fig:categories}
\end{figure}

The LeetCode website was employed as the source of data for this study. LeetCode is a platform designed for individuals seeking to enhance their programming abilities and/or prepare for job interviews in the field of computer science. The website offers a diverse range of programming problems that can be solved using various programming languages. The potential for mutual comparison and measurement is afforded by the competitive format. At the time of data collection (March 2024), the database contained 7,394 problems, which were assigned to different problem categories and levels of difficulty (\textit{easy}, \textit{medium} and \textit{hard}). Figure \ref{fig:categories} depicts the distribution of the problems across the categories, which reveals that in the top five categories the majority of problems are in the category \textit{Array}, followed by \textit{String}, \textit{Hash Table}, \textit{Math}, and \textit{Dynamic Programming}. It should be noted that a problem may be present in several categories, resulting in a total of 2,992 individual problems. However, between the start of the study (March 2024) and the end (August 2024), 188 new coding problems were provided by LeetCode that could not be included in this study.

The structure of a problem on LeetCode is always consistent and comprises the following elements:

\begin{enumerate}
    \item A description of the problem in text form.
    \item Examples with input, output and an explanation of how the output was obtained.
    \item Restrictions on parameter values.
\end{enumerate}
%
%
LeetCode provides several test cases and environments. Therefore, is a comprehensive and versatile coding problem platform, distinguished by a substantial number of problems, with a continuous expansion of the problem set, the capacity to evaluate solutions in a comprehensive manner, and a substantial corpus of handwritten solutions to the coding problems. These aspects are the primary determinants of the decision to collect samples via this platform, rather than utilising any of the briefly outlined benchmarks in Section \ref{sec:benchmarks}.

A Python script was developed with the objective of automating the entire sampling process, thereby reducing the time required for sample collection and enabling the process to be replicated as accurately as possible. LeetCode provides indirect support for web scraping through the design of the API. The platform employs GraphQL\footnote{GraphQL, \url{https://graphql.org/}}, a query language for communicating with the API, to retrieve data from the backend. Therefore, the requisite information was retrieved by means of a special GraphQL query, which was send using the Python library \textit{requests}\footnote{Requests, \url{https://pypi.org/project/requests/}}. This entailed setting specific variables, header and body parameters within the request. 

\subsubsection{Coding Problems}\label{subsubsec:CodingProblems}

In order to access the samples, it was first necessary to obtain the coding problems with the corresponding general information from LeetCode. Two stages were necessary to obtain the fundamental data: In the initial phase, data had to be requested for each category and problem, including the unique ID, the title, the acceptance rate, the difficulty level, the classification as either a premium question or a non-premium question, and the affiliation to other categories. This resulted in a dataset of 2,992 unique coding problems. In the subsequent stage, non-compliant coding problems were identified and rectified. Ultimately, a list was created for each category of problem, comprising the titles of the problems. More details on problem requirements are presented in the Appendix \ref{chap:app:pr}.

The final list of 2,321 programming problems serves as the foundation for the subsequent sample collection. As illustrated in Figure \ref{fig:research-process}, the initial step has been concluded and is now divided into two distinct paths. The first path delineates the gathering of human-crafted solution samples, while the second path outlines the process of collecting AI generated solution samples. Both paths are elaborated in the subsequent two chapters.

\subsubsection{Generate Solutions}
This section addresses the lower pipe of the flow chart of the sampling process, as illustrated in Figure \ref{fig:research-process}. The solutions were generated using the GenAI model \textit{GPT-4o} from OpenAI via API. 
In order to obtain valid generated solutions to the 2,321 programming problems, six steps were carried out in the Python script as part of the sampling process. The ensuing paragraphs will provide a concise overview of each of the following steps. Comprehensive details pertaining to these steps can be found in the Appendix \ref{chap:app:gs}.











\textbf{1. Request problem information from LeetCode.} The first step entails the retrieval of supplementary data from LeetCode for the purpose of formulating a prompt.

\begin{figure}
    \centering
    \subfloat[First Prompt]{
        \includegraphics[width=0.45\textwidth]{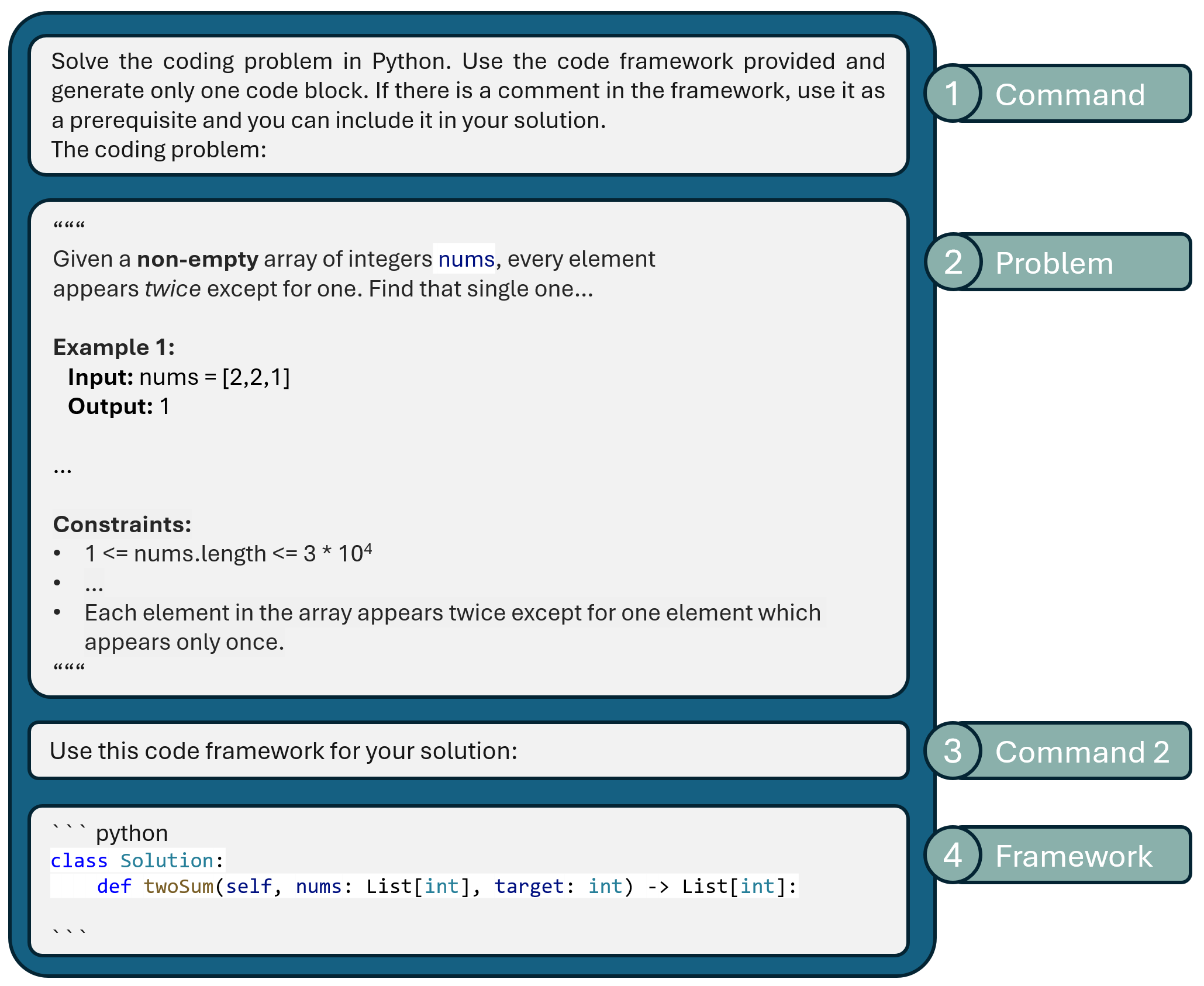}
        \label{fig:firstPrompt}
    }
    \hfill
    \subfloat[Prompt with Error\label{fig:errorPrompt}]{\includegraphics[width=0.45\textwidth]{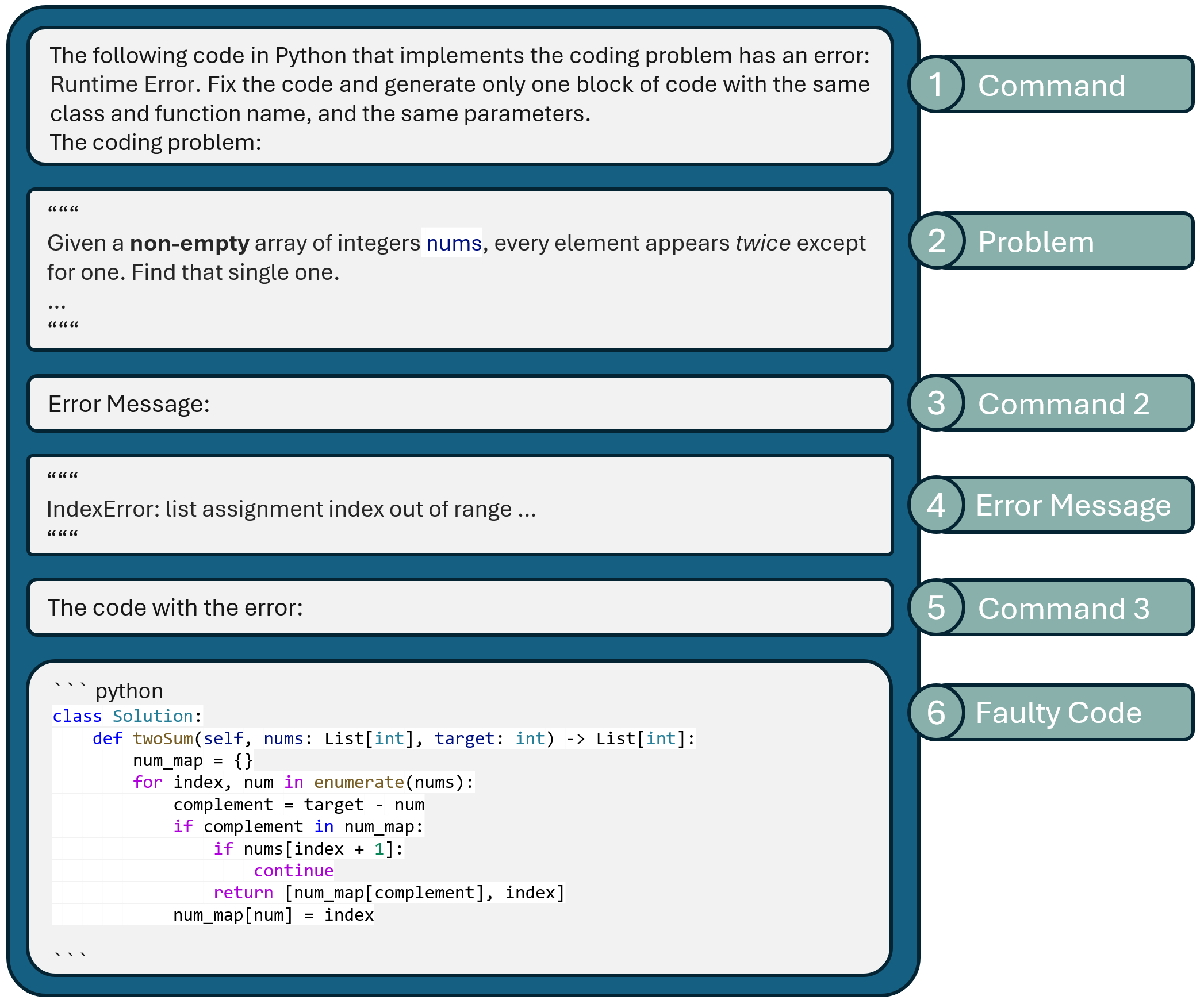}}
    \caption{Example prompts for OpenAI API}
    \label{fig:examplePrompt}
\end{figure}

\textbf{2. Assemble the input for the API call.} In the second step, a parallel iteration is conducted across all problem definitions, i.e. the description and the code framework, with the objective of assembling the prompt. The objective is not to consistently identify the optimal solution through the utilisation of sophisticated prompt designs; instead, it is to construct a query that closely approximates the actual scenario. An exemplar of the prompt can be found in Figure \ref{fig:firstPrompt}. The prompts were developed in accordance with the standards established by the model's creators\footnote{OpenAI Prompt Engineering, \url{https://platform.openai.com/docs/guides/prompt-engineering}}, as well as through a review of analogous studies for comparative purposes~\cite{tian2023chatgptultimateprogrammingassistant, Liu, LiuSQ}.

\textbf{3. Generate the code via OpenAI API.} In the third step of the process, interaction with the LLM was initiated. The \textit{gpt-4o-2024-05-13}\footnote{Model GPT-4o, \url{https://platform.openai.com/docs/models/gpt-4o}} model, representing the latest version of OpenAI at the time of the study, was employed. The model has a knowledge cutoff date of October 2023.

\textbf{4. Submit the code on LeetCode.} The generated code had to undergo a validation process prior to analysis. As part of the sampling process, the code was submitted to LeetCode for execution and testing using the test cases provided by LeetCode.

\textbf{5. Get information on submitted code.} The feedback on the submitted solutions was obtained from LeetCode.

\textbf{6. Evaluate submission.}\label{Step6}
In the event that the status was designated as ``Accepted'', the solution was stored and the generation of the solution for the subsequent problem commenced. Conversely, in the event of an error occurring during the execution of the code or other issues arising, a multi-round fixing approach with a maximum of five attempts was employed, similar to~\cite{Liu}, see Appendix \ref{chap:app:gs} for details. For the retries, the prompt was assembled accordingly to the error. An example is provided in Figure \ref{fig:errorPrompt}.

Upon completion of Steps 1 - 6 for all 2,321 programming problems, 3,676 prompts were created and Python files comprising 110,819 lines of code were generated. Consequently, the validation process conducted by LeetCode permitted the acceptance of 2,086 of the generated solutions, which represents a total of 89.88\% of the overall generated solutions. Of the accepted problems, a total of 53,509 lines of generated code were identified, of which 35,122 lines were pure code, devoid of blank lines or comments.

\subsubsection{User Solutions}
This section addresses the upper pipe of the flow chart of the sampling process, as illustrated in Figure \ref{fig:research-process}. In order to obtain the handwritten samples from the users, a total of four steps were required. 
The individual steps are described below. The sampling process commenced with the initial list of coding problem titles from Section \ref{subsubsec:CodingProblems}, with an iteration from top to bottom. A more comprehensive description of each of the following steps is provided in the Appendix \ref{chap:appendix:us}.


    





            

\textbf{1. Query list of all community posts.}
Once a solution to the problem has been successfully developed and all test cases have been completed successfully on LeetCode, users are given the option of making the solution publicly available as a community post. This data contains an post ID, user text (Markdown), title, tags, evaluation by other users (upvotes), and comments. Therfore, the first step encompasses to query the list of all communtiy post IDs per coding problem.

\textbf{2. Query each community post.}
By iterating over the list of posts from the previous step, the content of the post was requested using a query with the ID of the post.

\textbf{3. Extract code snippets from post.}
The third stage of the process involved the extraction of all code blocks from the Markdown document and their subsequent assignment to a specific programming language.

\textbf{4. Validate user solution.}
In the final stage of the process, the non-valid code snippets were removed, and the Python code solutions were consolidated into a single solution per post and problem. In order to achieve this, a local evaluation of the code was conducted. In some cases, multiple code snippets with the same identified programming language were present within a single post. In this manner, the code was frequently assembled in a step-by-step manner, with the final solution typically presented as the concluding element. Consequently, the code snippets were evaluated in a sequential manner, from the latter to the former, with the objective of identifying any valid code. As the objective was to examine solely Python code snippets, a comprehensive analysis of the code was conducted in cases where Python had been previously assigned to the snippet.

A total of 278,397 posts were collected for analysis; of this number, 70,261 posts were identified as containing at least one Python code snippet. Following the removal of non-valid code, 57,238 valid Python solutions were identified, representing 81.5\% of the total. The code snippets encompass 1,258,278 lines of code and 923,452 pure lines of code, excluding blank lines and comments. The overall statistics for the collected samples are presented in Table \ref{tab:overallSamples}. Consequently, this study yielded a total of 59,324 Python code snippets, comprising 1,311,787 source lines of code and 958,574 pure lines of code.

\begin{table}[h]
\centering
\begin{tabular*}{\textwidth}{@{\extracolsep{\fill}}lrrrr}
\toprule
 & \textbf{Solutions} & \textbf{Valid Solutions} & \textbf{SLOC} & \textbf{LOC} \\
\midrule
\textbf{Generated} & 2,321 & 2,086 & 53,509 & 35,122 \\
\textbf{User Created} & 70,261 & 57,238  & 1,258,278 & 923,452 \\
\midrule
\textbf{Total} & 72,582 & 59,324 & 1,311,787 & 958,574 \\
\bottomrule
\end{tabular*}
\caption{Statistics of collected samples}
\label{tab:overallSamples}
\end{table}

\subsection{Data Mining}\label{sec:datamining}
The following section is dedicated to the second section of Figure \ref{fig:research-process}, data collection. A total of four metrics were used for the first research questions, including only quantitative data. A Python script was developed for data collection, which automates the process. In order to obtain the metrics, two data sources were employed: the first was the platform LeetCode itself with GraphQL queries, while the second was the static code analysis tool SonarQube and its API. SonarQube is a tool that performs a local static analysis of the code using the SonarScanner, whereby the results are evaluated and visualised on a SonarQube server. It is important to note that in a large-scale study, the standard SQLite\footnote{SQLite, \url{https://www.sqlite.org/}} database of SonarQube is unable to cope with the sheer volume of entries and almost parallel accesses. Consequently, a PostgreSQL\footnote{PostgreSQL, \url{https://www.postgresql.org/}} database was introduced as a replacement, offering superior performance and resilience to the aforementioned issues.

\subsubsection{Code Quality}\label{sec:codeQuali}
The initial stage of the data collection and analysis process entailed an examination of the code smells. The analysis of code smells enables the identification of poor programming patterns, which can serve as an indicator of low code quality. It is vital that these code smells are identified and eradicated in a timely manner; otherwise, they can result in a build-up of technical debt~\cite{Palomba} and the emergence of bugs~\cite{van2012assuring}. Consequently, the presence of code smells has an impact on the maintainability of the code~\cite{yamashita2013good}, as also reflected by SonarQube. The code smells were quantified using SonarQube, with a local analysis conducted for each of the 59,324 Python files, including the samples of the user and the generated samples. As part of the analytical process, 152 rules were applied to the Pyton source code, for instance, the rule designated as \textit{``Builtins should not be shadowed by local variables''} or \textit{``Functions and methods should not be empty''}. The rules are classified (e.g. \textit{Consistency}, \textit{Intentionality}, \textit{Responsibility}) and presented in accordance with their level of impact on the maintainability (\textit{low}, \textit{medium}, \textit{high}). The local analysis and the determination of the metrics were conducted in three steps, which were implemented in the Python script:

\begin{enumerate}
    \item A unique token was generated for each code file, as well as a unique title (structured as \textit{\{postVotes\}\_\{postId\}}).
    \item The command to initiate the local analysis was executed with the requisite information pertaining to the unique token and title.
    \item Following the processing of the analysis on the SonarQube server, the metric could be retrieved via a request to the API with the unique title.
\end{enumerate}

As the number of code smells is not a meaningful variable due to its dependence on project size, the code quality for individual Python solutions was operationalised using the \textit{number of code smells per LOC}. To this end, SonarQube was employed to ascertain \textit{ncloc}, which encompasses the number of lines of code, excluding comments and empty lines.

\subsubsection{Code Understandability}\label{sec:codeUnder}
A further, more specific type of code quality is the degree of understandability of the code for developers. The assessment of cognitive complexity in relation to the understandability of a code snippet is emerging as a promising metric~\cite{Marvin}, having gained increasing prominence in recent times. Cognitive complexity is a score based on the structural and flow characteristics of the source code. The evaluation is based on three rules, which are as follows:

\begin{enumerate}
    \item The score is increased with each break in the linear flow of the code. This encompasses loop structures, conditionals, switches and catch statements, sequences of logical operators, recursion and jumps to labels.
    \item Additionally, the score is augmented with each increment of nesting. To illustrate this, the following example: In the event that a condition exists within a loop, the nesting is increased by a factor of one, resulting in an initial score of two. Consequently, an additional nesting would result in an increase of the score by two for the nesting, and so forth.
    \item In evaluating the structures, any readable, shorthand structures are disregarded. Consequently, structures that permit the consolidation of multiple statements into a concise, comprehensible format are not considered. To illustrate, an increase in the score is not observed when null coalescing operators are employed, that is, when the ``?'' symbol is utilised to ascertain whether the variable possesses the value null or is undefined.~\cite{Campbell}
\end{enumerate}

Consequently, a higher score indicates that the code snippet is more challenging to understand.

The cognitive complexity score is calculated by SonarQube during the analysis and was requested via the API. In a manner analogous to the code quality, the code understandability was operationalised using the \textit{cognitive complexity score per line of code}.

\subsubsection{Time Behaviour}
In the ISO 25010 definition\footnote{ISO 25010 software quality, \url{https://iso25000.com/index.php/en/iso-25000-standards/iso-25010/59-performance-efficiency}}, the time behaviour is treated in terms of performance efficiency, which allows a direct influence on software quality to be derived. Consequently, the runtime of the code during execution is quantified. Nevertheless, the assessment is not conducted at the local level; rather, it is provided by LeetCode subsequent to the submission of a solution. The data is presented in millisecond units. However, as emphasised by the authors Choudhuri et al.~\cite{Choudhuri}, the metrics offered by LeetCode do not consistently align closely with the times recorded locally. On average, a slight correlation can be observed; in some cases, though, a high correlation can also be seen. Furthermore, a higher degree of variance was observed. These factors must be taken into account when analysing the data, and the results must be interpreted with caution. In order to ensure comparability of performance with that of other LeetCode users, the \textit{runtime rank} provided by LeetCode was considered. The value is expressed as a percentile and indicates the proportion of users whose runtime speed is slower than that of the submitted solution. For example, a value of 60 indicates that the submitted solution is faster than 60\% of other users' solutions in terms of runtime. The metrics of the rank were collected via a GraphQL query after generating the solution for the individual programming problems and uploading it to LeetCode. 

\subsubsection{Resource Utilisation}
Resource utilisation, like time behaviour, is included under performance efficiency in the ISO for software quality. The measurement of resource utilisation is based on the memory usage in megabytes required during code execution. In this context, the metrics employed by LeetCode are utilised, with a comparison of the generated solution with the users of LeetCode facilitated by the \textit{memory usage rank}.

\section{Analysis and Results}
The research question assesses the large-scale data mining aspect of the study, comprising four hypotheses that were subjected to a range of hypothesis tests. The specific test method is detailed for each hypothesis in the corresponding section. For enhanced clarity, a box plot is provided for each hypothesis. Prior to the presentation of the hypothesis results, an overview of the collected data is provided in Section \ref{sec:generalData}. Subsequently, the hypotheses are evaluated in Sections \ref{sec:H11} - \ref{sec:H14}.

\subsection{General Data}\label{sec:generalData}
Given the considerable volume of data obtained through web scraping and data mining, it is imperative to undertake an initial overview of the samples in order to evaluate the hypotheses. The following section presents a statistical overview of the samples.

\begin{table}[h]
    \centering
    \begin{tabularx}{\textwidth}{>{\raggedright\arraybackslash}p{3cm} *{6}{>{\raggedleft\arraybackslash}X} c}
    \toprule
    & \multicolumn{2}{c}{\textbf{Easy}} & \multicolumn{2}{c}{\textbf{Medium}} & \multicolumn{2}{c}{\textbf{Hard}} \\
    \cmidrule(rr){2-3} \cmidrule(rr){4-5} \cmidrule(rr){6-7}
    \textbf{Problems} & \multicolumn{1}{c}{\textbf{\#}} & \multicolumn{1}{c}{\textbf{\%}} & \multicolumn{1}{c}{\textbf{\#}} & \multicolumn{1}{c}{\textbf{\%}} & \multicolumn{1}{c}{\textbf{\#}} & \multicolumn{1}{c}{\textbf{\%}} & \multicolumn{1}{c}{\textbf{Total}}\\
    \midrule
    & \\[\dimexpr-\normalbaselineskip+2pt]
    GPT-4o solved & 573 & 27.47 & 1,125 & 53.93 & 388 & 18.60 & 2,086\\
    & \\[\dimexpr-\normalbaselineskip+2pt]
    User solved & 577 & 24.94 & 1,207 & 52.16 & 530 & 22.90 & 2,314 \\
    & \\[\dimexpr-\normalbaselineskip+2pt]
    Total & 579 & 24.95 & 1,209 & 52.09 & 533 & 22.96 & 2,321 \\
    \bottomrule
    \end{tabularx}
    \caption{Overview of problems with valid generated solutions by GPT-4o and valid created solutions by LeetCode user}
    \label{tab:ProblemOverview}
    \vspace{2mm}
    \footnotesize{\# indicates the number of solutions and \% indicates the percentage of the difficulty category in relation to the total}
\end{table}

Table \ref{tab:ProblemOverview} provides an overview of the number of coding problems for which samples have been collected. It is essential to categorise the number of problems according to difficulty level, as this has implications for the methodology of the statistical analysis. In total, 2,321 coding problems were sourced from LeetCode. However, a solution in Python created by an user could not be recorded for every problem, resulting in the final number of 2,314 problems for which a valid user solution has been provided. Consequently, seven problems remain without a solution in Python. A total of 2,086 valid solutions were generated for the programming problems by GPT-4o, while 235 generated solutions of the total 2,321 were invalid. A solution deemed valid could be achieved within a maximum generation of five attempts. This yields a valid solution rate of 89.88\% for the generation by the GPT-4o model. As evidenced by the data presented in the table, it appears that GPT-4o encounters greater challenges in generating valid solutions as the difficulty level of the problems increases. Consequently, the distribution of problem difficulty levels is uneven, with a notable discrepancy in the proportion of \textit{Hard} problems. Specifically, 18.60\% of the problems for which a valid solution was generated by GPT-4o are classified as \textit{Hard}, compared to 22.90\% of the problems for which a valid solution is available from the developers on LeetCode.

\begin{table}[h]
    \centering
    \begin{tabularx}{\textwidth}{p{2cm} *{8}{>{\raggedleft\arraybackslash}X}}
    \toprule
    & \multicolumn{2}{c}{\textbf{Easy}} & \multicolumn{2}{c}{\textbf{Medium}} & \multicolumn{2}{c}{\textbf{Hard}} & \multicolumn{2}{c}{\textbf{Total}}\\
    \cmidrule(rr){2-3} \cmidrule(rr){4-5} \cmidrule(rr){6-7} \cmidrule(rr){8-9}
    \textbf{Problems} & \multicolumn{1}{c}{\textbf{M}} & \multicolumn{1}{c}{\textbf{Mdn}} & \multicolumn{1}{c}{\textbf{M}} & \multicolumn{1}{c}{\textbf{Mdn}} & \multicolumn{1}{c}{\textbf{M}} & \multicolumn{1}{c}{\textbf{Mdn}} & \multicolumn{1}{c}{\textbf{M}} & \multicolumn{1}{c}{\textbf{Mdn}}\\
    \midrule
    & \\[\dimexpr-\normalbaselineskip+2pt]
    GPT-4o solved & 64.90 & 65.50 & 54.00 & 53.30 & 47.09 & 46.00 & 55.71 & 55.10 \\
    & \\[\dimexpr-\normalbaselineskip+2pt]
    User solved & 64.71 & 65.40 & 52.93 & 52.20 & 44.42 & 43.20 & 53.92 & 53.05\\
    & \\[\dimexpr-\normalbaselineskip+2pt]
    Total & 64.72 & 65.40 & 52.91 & 52.20 & 44.41 & 43.20 & 53.90 & 53.00\\
    \bottomrule
    \end{tabularx}
    \caption{Overview of the acceptance rate of the problems}
    \label{tab:ProblemsAccRate}
    \vspace{2mm}
    \footnotesize{Values are presented in \%. M indicates mean and Mdn denotes the median of the acceptance rate}
\end{table}

Table \ref{tab:ProblemsAccRate} illustrates the corresponding mean and median acceptance rates for the aforementioned coding problems, as determined by LeetCode. This key figure provides information regarding the proportion of valid solutions submitted on LeetCode. With regard to all problems, the mean shows a rate of 53.90\%, indicating that more than every second solution submitted to LeetCode is valid. As anticipated, the rate declines with increasing difficulty. It is noteworthy that for problems for which valid generated solutions are available, the acceptance rate for each difficulty level is higher than for problems for which solutions from users are available. With regard to the median, a difference of 2.80\% can be observed at the \textit{Hard}s difficulty level. In general, this indicates that GPT-4o encountered greater difficulty in finding solutions for problems with a lower acceptance rate.

\begin{table}[h]
    \centering
    \begin{tabularx}{\textwidth}{>{\raggedright\arraybackslash}p{2.5cm} *{6}{>{\raggedleft\arraybackslash}X} c}
    \toprule
    & \multicolumn{2}{c}{\textbf{Easy}} & \multicolumn{2}{c}{\textbf{Medium}} & \multicolumn{2}{c}{\textbf{Hard}} \\
    \cmidrule(rr){2-3} \cmidrule(rr){4-5} \cmidrule(rr){6-7}
    \textbf{Samples} & \multicolumn{1}{c}{\textbf{\#}} & \multicolumn{1}{c}{\textbf{\%}} & \multicolumn{1}{c}{\textbf{\#}} & \multicolumn{1}{c}{\textbf{\%}} & \multicolumn{1}{c}{\textbf{\#}} & \multicolumn{1}{c}{\textbf{\%}} & \multicolumn{1}{c}{\textbf{Total}}\\
    \midrule
    & \\[\dimexpr-\normalbaselineskip+2pt]
    GPT-4o generated & 573 & 27.47 & 1,125 & 53.93 & 388 & 18.60 & 2,086\\
    & \\[\dimexpr-\normalbaselineskip+2pt]
    User created & 19,401 & 33.90 & 29,885 & 52.21 & 7,952 & 13.89 & 57,238 \\
    \bottomrule
    \end{tabularx}
    \caption{Overview of solutions created by GPT-4o and LeetCode user}
    \label{tab:SolutionsSamples}
    \vspace{2mm}
    \footnotesize{\# indicates the number of solutions and \% indicates the percentage of the difficulty category in relation to the total.}
\end{table}

In order to investigate the coding problems presented in Table \ref{tab:ProblemOverview}, the corresponding samples were collected. In total, 2,086 generated solutions (one valid solution per problem) and 57,238 solutions from users were collated. Table \ref{tab:SolutionsSamples} depicts the respective proportions of the collected solutions to the difficulty levels, in addition to the problem count. Once again, divergent proportions of difficulty emerge, particularly in the \textit{Easy} and \textit{Hard} categories, where the discrepancy between generated and handwritten samples is noteworthy, at 6.43\% and 4.71\%, respectively. In regard to the hypotheses that involve a direct comparison between the user samples and the generated samples (H$_{1}^{1}$ and H$_{1}^{2}$), the following actions were undertaken: Given the disparate sample sizes for each difficulty level and the varying acceptance rates for problems in different difficulty categories, only the solutions matching the intersection of problems from Table \ref{tab:ProblemOverview} of the generated samples and the user samples were included in the subsequent analysis. Furthermore, the mean was calculated for the user solutions for each problem in order to facilitate a direct comparison between the generated solutions and the average developer. This approach avoids any potential distortion of the influence of the difficulty of the problems and ensures a fair comparison in terms of the distribution of user solutions to the problems.

\begin{table}[h]
    \centering
    \begin{tabularx}{\textwidth}{>{\raggedright\arraybackslash}p{2.3cm} *{8}{>{\raggedleft\arraybackslash}X} >{\raggedleft\arraybackslash}X}
    \toprule
    & \multicolumn{2}{c}{\textbf{Easy}} & \multicolumn{2}{c}{\textbf{Medium}} & \multicolumn{2}{c}{\textbf{Hard}} & \multicolumn{2}{c}{\textbf{Total}} \\
    \cmidrule(rr){2-3} \cmidrule(rr){4-5} \cmidrule(rr){6-7} \cmidrule(rr){8-9}
    \multicolumn{1}{l}{\textbf{Problems}} & \multicolumn{1}{c}{\textbf{\#}} & \multicolumn{1}{c}{\textbf{\%}} & \multicolumn{1}{c}{\textbf{\#}} & \multicolumn{1}{c}{\textbf{\%}} & \multicolumn{1}{c}{\textbf{\#}} & \multicolumn{1}{c}{\textbf{\%}} & \multicolumn{1}{c}{\textbf{\#}} & \multicolumn{1}{c}{\textbf{GenAI Solv.}}\\
    \midrule
    before-problems & 527 & 24.92 & 1,108 & 52.38 & 480 & 22.67 & 2,115 & 1,975\\
    after-problems & 52 & 25.24 & 101 & 49.03 & 53 & 25.73 & 206 & 107\\
    \midrule
    Total & 579 & 24.95 & 1,209 & 52.09 & 533 & 22.96  & 2,321 & 2,086\\
    \bottomrule
    \end{tabularx}
    \caption{Overview of problems introduced before and after October 2023}
    \label{tab:BefAftProblems}
    \vspace{2mm}
    \footnotesize{\# indicates the number of solutions and \% indicates the percentage of the difficulty category in relation to the total. GenAI Solv. indicates the number of solved problems by the GenAI}
\end{table}

In light of the fact that certain studies only incorporate problems that are not included in the training data set of the GenAI, the problems were classified in Table \ref{tab:BefAftProblems} into two distinct categories: those that emerged prior to the cutoff date of the training data set, which are designated as \textit{before-problems}, and those that originated subsequent to the cutoff date, which are designated as \textit{after-problems}. It is not feasible to determine the exact date of the cutoff. The only available information is that the training data is up until October 2023\footnote{Cutoff date training data, \url{https://platform.openai.com/docs/models/gpt-4o}}. Accordingly, a cutoff date of 1 October 2023 was established. The table additionally illustrates the number and ratio per difficulty category, the total number of problems included in the before and after categories, and the number of problems solved by the GenAI. A total of 2,115 before-problems and 206 after-problems were identified. It is noteworthy that the GenAI only generated a valid solution for just over half of the after-problems, 107 in total. The solution rate for the problems contained in the training dataset was 93.38\%, which corresponds to 1,975 out of 2,115 problems.

\begin{figure}[ht]
    \centering
    \includegraphics[width=\columnwidth]{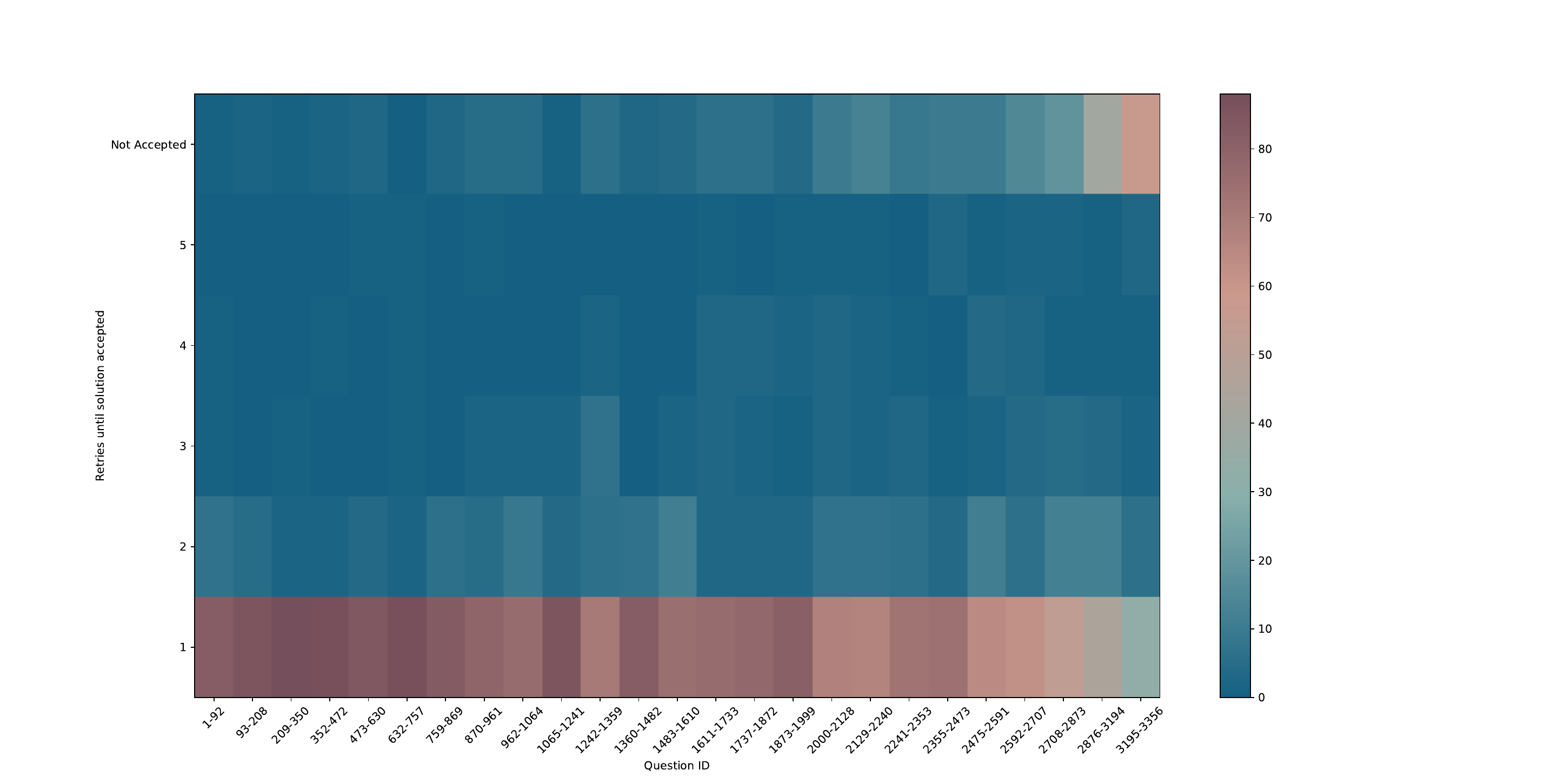}
    \vspace{-15pt}
    \caption{Heatmap of questions by id and retries until the solution is accepted}
    \label{fig:heatmap-id-retries}
\end{figure}

Figure \ref{fig:heatmap-id-retries} depicts a heat map that illustrates the problem question ID (which is incremented for new problems on LeetCode) on the x-axis and the number of attempts until the GenAI has generated a valid solution on the y-axis. The maximum value on the y-axis represents the number of problems for which a valid solution could not be generated. Question ID 2,873 represents the cutoff date for the training data, thus marking the final inclusion in the before-problems category. It is evident from the heat map that the GenAI encountered increasing challenges in solving the problems as it approached the cutoff date. The heatmap clearly demonstrates that the rate of problem-solving is notable lower after the cutoff, a finding that is also evident in Table \ref{tab:BefAftProblems}. However, given that the sample size for the after-problems is relatively small (107), and the objective is to consider the entire dataset, the subsequent analyses only suggest discrepancies between the tests of the entire dataset and the after-problems. Otherwise, the tests are always conducted with the entire dataset.

\subsection{H$_{1}^{1}$ GenAI produces better code quality than developers on LeetCode.}\label{sec:H11}
The initial hypothesis concerning code quality was assessed by comparing the operationalised metric, \textit{number of code smells per kLOC}, of the generated code snippets by GPT-4o with that of the code snippets written by the users. In order to facilitate a more straightforward and readily comprehensible understanding, the metric was extrapolated to code smells per thousand lines of code. The data pertinent to the initial hypothesis are presented in Table \ref{tab:CodeSmellsLeet}.

\begin{table}[h]
    \centering
    \begin{tabular}{lllllll}
        \toprule
        \textbf{Samples} & \multicolumn{1}{c}{\textbf{Problems}} & \multicolumn{1}{c}{\textbf{Solutions}} & \multicolumn{1}{c}{\textbf{Code Smells}} & \multicolumn{1}{c}{\textbf{LOC}} & \multicolumn{1}{c}{\textbf{M}} & \multicolumn{1}{c}{\textbf{Mdn}}\\
    \midrule
    GPT-4o solutions & \multicolumn{1}{r}{2,082} & \multicolumn{1}{r}{2,082} & \multicolumn{1}{r}{2,906} & \multicolumn{1}{r}{35,029} & \multicolumn{1}{r}{94.23} & \multicolumn{1}{r}{76.92}\\
    User solutions & \multicolumn{1}{r}{2,082} & \multicolumn{1}{r}{55,392} & \multicolumn{1}{r}{89,825} & \multicolumn{1}{r}{876,748} &\multicolumn{1}{r}{115.94}  & \multicolumn{1}{r}{104.17}\\
    \bottomrule
    \end{tabular}
    \caption{Code quality results}
    \label{tab:CodeSmellsLeet}
    \vspace{2mm}
    \footnotesize{Mean (M) and median (Mdn) are indicated as number of code smells per kLOC}
\end{table}

\begin{figure}[ht]
    \centering
    \includegraphics[width=\columnwidth]{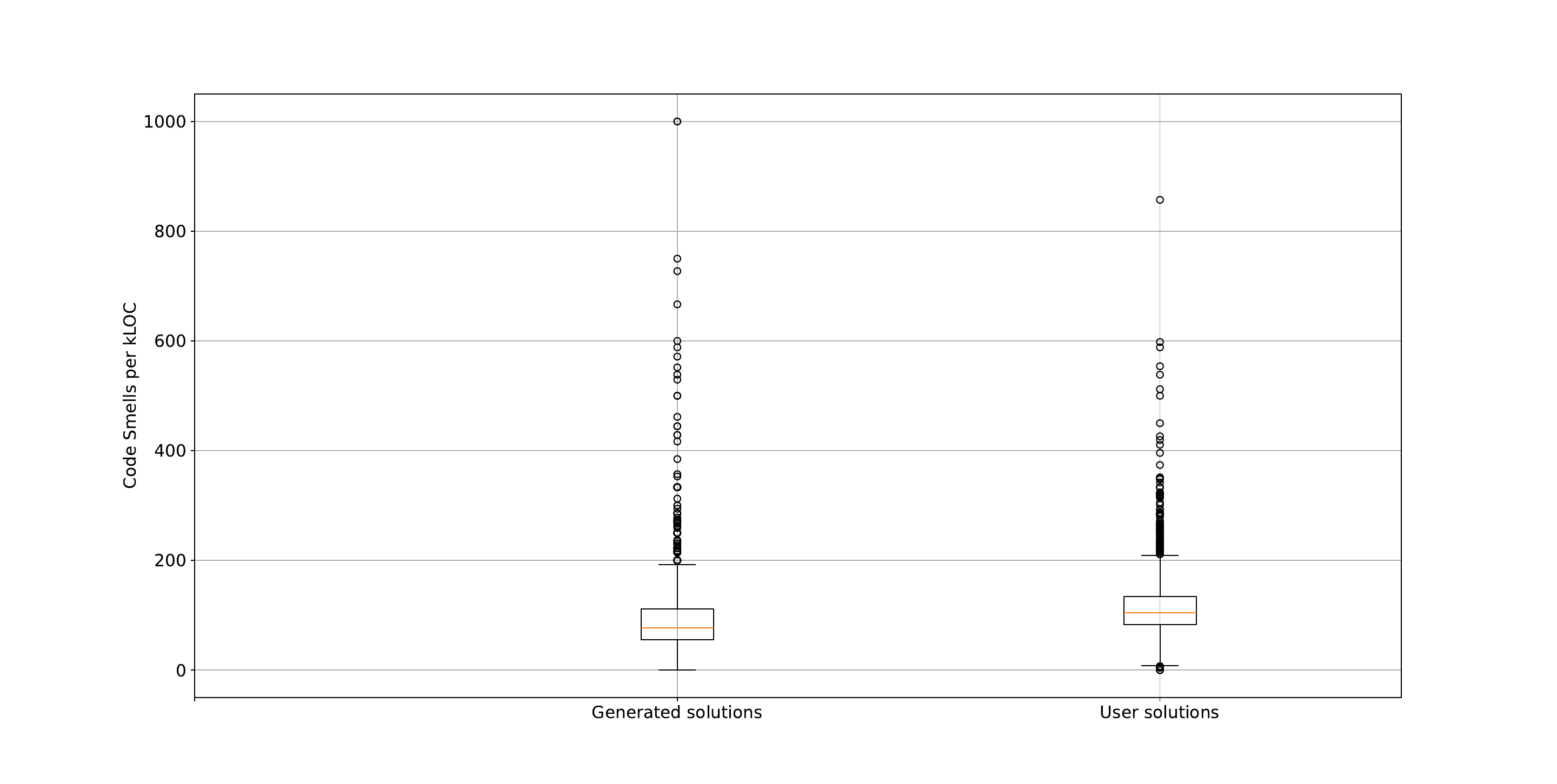}
    \vspace{-15pt}
    \caption{Boxplot of code smells per kLOC of the generated solutions and the user solutions}
    \label{fig:bxpl_code_smells_leet}
\end{figure}

The generated and human-written samples were subjected to analysis for a total of 2,082 programming problems. In the user code snippets, 89,825 instances of code smells were identified, which relate to 55,392 solutions and 876,748 lines of code. However, the average value per problem was employed for the user solutions. In the context of solution generation, a single solution was created for each problem. This resulted in a total of 2,082 solutions and 35,029 lines of code, which were then analysed to identify 2,906 instances of code smells. Figure \ref{fig:bxpl_code_smells_leet} illustrates the descriptive statistics of the distribution of the two data sets of code smells per kLOC, represented by box plots. The median value of the generated solutions is less than that of the solutions created by developers, with 76.92 code smells per kLOC for the generated solutions and 104.17 for the user solutions. The analysis indicates that the code quality of the solutions written by users is 27.25 code smells per kLOC worse than that of the generated code. It is, however, noteworthy that the box plots of the two groups exhibit a slight offset, with the data sets of the generated solutions displaying slightly lower values. Nevertheless, there is a greater prevalence of outliers in the generated code, which brings the mean closer to that of the solutions written by humans.

The already lower median of the generated sample provides support for the correct assumption made in formulating the alternative hypothesis. Given the non-normal distribution of the data set, the Mann-Whitney U hypothesis test was conducted. The p-value of $2.13 \times 10^{-89}$ is less than the adjusted significance level of $1.25 \times 10^{-02}$, indicating that the null hypothesis can be rejected and the alternative hypothesis accepted. Additionally, Cohen's d indicates a value of 0.65, suggesting a medium-sized effect between the difference in distribution of the generated and human-written samples.

\textbf{Result for H$_{1}^{1}$:} Code solutions generated by GPT-4o exhibit a significantly lower prevalence of code smells in comparison to those developed by human coders on LeetCode. Consequently, the quality of the generated code by GPT-4o is superior for LeetCode problems.

\subsection{H$_{1}^{2}$ GenAI produces better code understandability than developers on LeetCode.}
The second hypothesis seeks to assess the understandability of the code between the two sample groups using the operationalised metric \textit{cognitive complexity score per kLOC}. Similarly, the cognitive complexity score per line of code was extrapolated to a thousand lines of code, in a manner analogous to the approach taken with regard to the code smells. The data collected for this purpose is presented in Table \ref{tab:CogComplLeet}.

\begin{table}[h]
    \centering
    \begin{tabular}{lllllll}
        \toprule
        \textbf{Samples} & \multicolumn{1}{c}{\textbf{Problems}} & \multicolumn{1}{c}{\textbf{Solutions}} & \multicolumn{1}{c}{\textbf{CC Score}} & \multicolumn{1}{c}{\textbf{LOC}} & \multicolumn{1}{c}{\textbf{M}} & \multicolumn{1}{c}{\textbf{Mdn}}\\
    \midrule
    GPT-4o solutions & \multicolumn{1}{r}{2,082} & \multicolumn{1}{r}{2,082}  & \multicolumn{1}{r}{15,774} & \multicolumn{1}{r}{35,029} & \multicolumn{1}{r}{405.18} & \multicolumn{1}{r}{375.00}\\
    User solutions & \multicolumn{1}{r}{2,082} & \multicolumn{1}{r}{55,392} & \multicolumn{1}{r}{383,560} & \multicolumn{1}{r}{876,748} &\multicolumn{1}{r}{423.42}  & \multicolumn{1}{r}{405.29}\\
    \bottomrule
    \end{tabular}
    \caption{Code understandability results}
    \label{tab:CogComplLeet}
    \vspace{2mm}
    \footnotesize{Mean (M) and median (Mdn) are indicated as a cognitive complexity (CC) score per kLOC}
\end{table}

\begin{figure}[ht]
    \centering
    \includegraphics[width=\columnwidth]{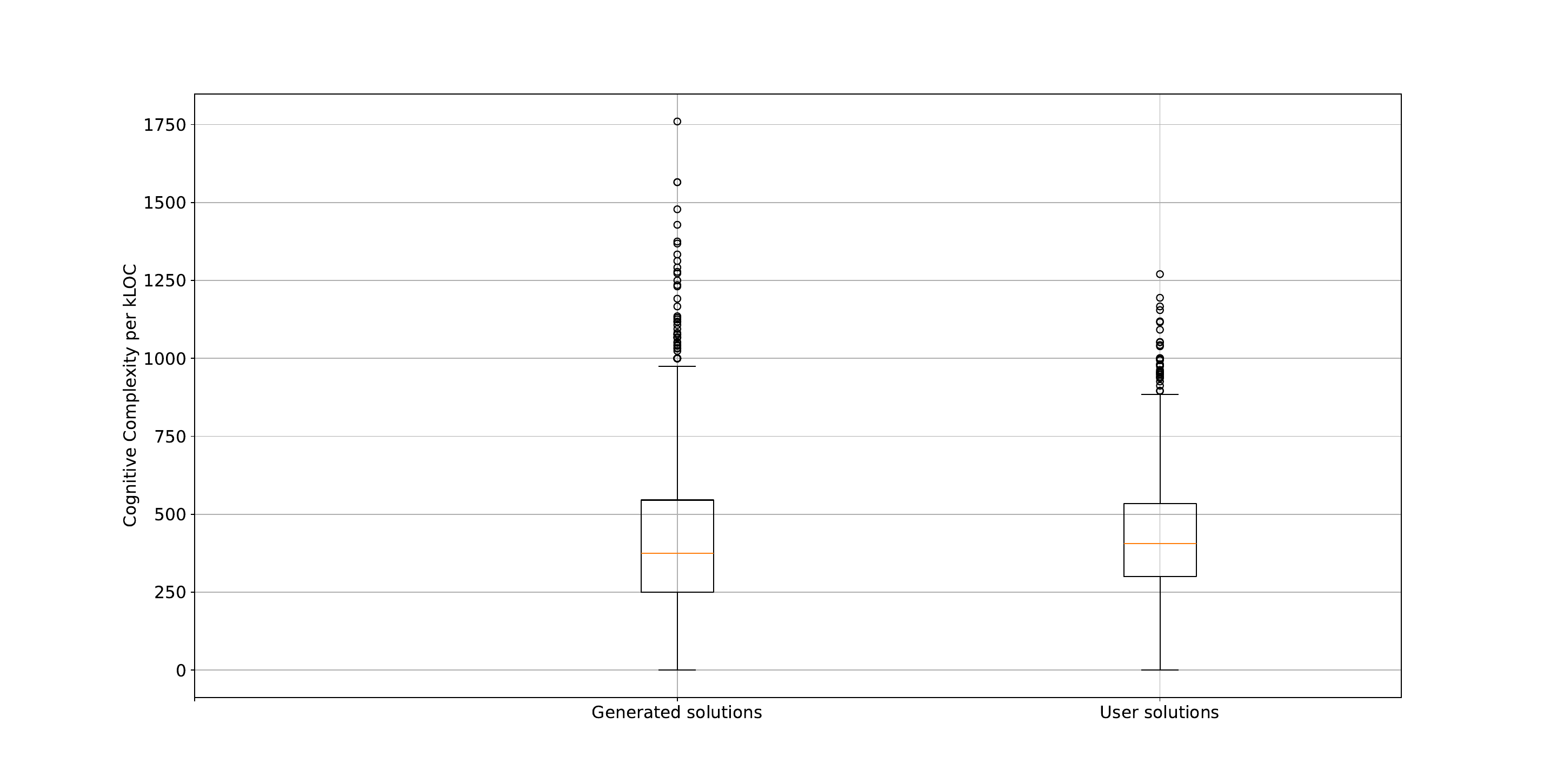}
    \vspace{-15pt}
    \caption{Boxplot of cognitive complexity per kLOC of the generated solutions and the user solutions}
    \label{fig:bxpl_cog_comple_leet}
\end{figure}

The same set of the 2,082 problems was employed in this hypothesis as in the previous one. A total of 35,029 lines of code and a cognitive complexity score of 15,774 were determined for the 2,082 solutions generated. For the 55,392 solutions provided by LeetCode users, 876,748 lines of code were measured with a cognitive complexity score of 383,560. Once again, the average cognitive complexity score per problem was calculated for the user solutions. As illustrated in Figure \ref{fig:bxpl_cog_comple_leet}, the box plots offer a compelling representation of the data distribution. It is evident that the distribution of the generated solutions is more dispersed than that of the solutions created by humans. Furthermore, the outliers are considerably larger, which is likely attributable to the average per problem for solutions authored by users. Nevertheless, the interquartile range (IQR) and median for the generated solutions are lower. When evaluated on a per kLOC basis, the median of the generated code, at 375, exhibits a slightly lower cognitive complexity score than that of user-written code, at 405.29, representing a difference of 30.29.

The lower median offers an insight into the result of the hypothesis test. Given that the datasets in question are not normally distributed, the Mann-Whitney U test was employed in this instance. The p-value of $4.01 \times 10^{-06}$ is less than the significance level of $1.25 \times 10^{-02}$, indicating that the null hypothesis can be rejected and the alternative hypothesis accepted. With regard to the effect size of Cohen's d, it can be observed that this has a small effect with \texttt{d = 0.14}. With regard to this hypothesis, it should be noted that the test result for the \textit{after-problems} with a p-value of 0.43 would yield a different result.

\textbf{Result for H$_{1}^{2}$:} Code solutions generated by GPT-4o exhibit a significantly lower cognitive complexity score in comparison to those developed by human coders on LeetCode. Consequently, the code understandability of the generated code by GPT-4o is superior for LeetCode problems.

\subsection{H$_{1}^{3}$ GenAI produces code that utilises less resources than developers on LeetCode.}
This hypothesis compares the solutions devised by LeetCode users with the generated solutions in terms of the utilisation of resources during the execution of the solution. The comparison is based on the \textit{memory usage rank} on LeetCode. An overview of the pertinent information can be found in Table \ref{tab:RessourcesTotal}, including the mean and median memory usage rank.

\begin{table}[h]
    \centering
    \begin{tabular}{lllllll}
        \toprule
        \textbf{Samples} & \multicolumn{1}{c}{\textbf{Problems}} & \multicolumn{1}{c}{\textbf{Mean}} & \multicolumn{1}{c}{\textbf{Median}}\\
    \midrule
    GPT-4o solutions & \multicolumn{1}{r}{2,086} & \multicolumn{1}{r}{49.02} & \multicolumn{1}{r}{48.16}\\
    \bottomrule
    \end{tabular}
    \caption{Memory usage rank of generated soltuions on LeetCode}
    \label{tab:RessourcesTotal}
    \vspace{2mm}
    \footnotesize{Mean and median are presented in \%}
\end{table}

\begin{table}[h]
    \centering
    \begin{tabular}{lllllll}
        \toprule
        \textbf{Samples} & \multicolumn{1}{c}{\textbf{Problems}} & \multicolumn{1}{c}{\textbf{Mean}} & \multicolumn{1}{c}{\textbf{Median}}\\
    \midrule
    GPT-4o solutions & \multicolumn{1}{r}{388} & \multicolumn{1}{r}{52.04} & \multicolumn{1}{r}{53.87}\\
    \bottomrule
    \end{tabular}
    \caption{Memory usage rank of generated soltuions on LeetCode for \textit{Hard} problems}
    \label{tab:RessourcesHard}
    \vspace{2mm}
    \footnotesize{Mean and median are presented in \%}
\end{table}

\begin{figure}[ht]
    \centering
    \includegraphics[width=\columnwidth]{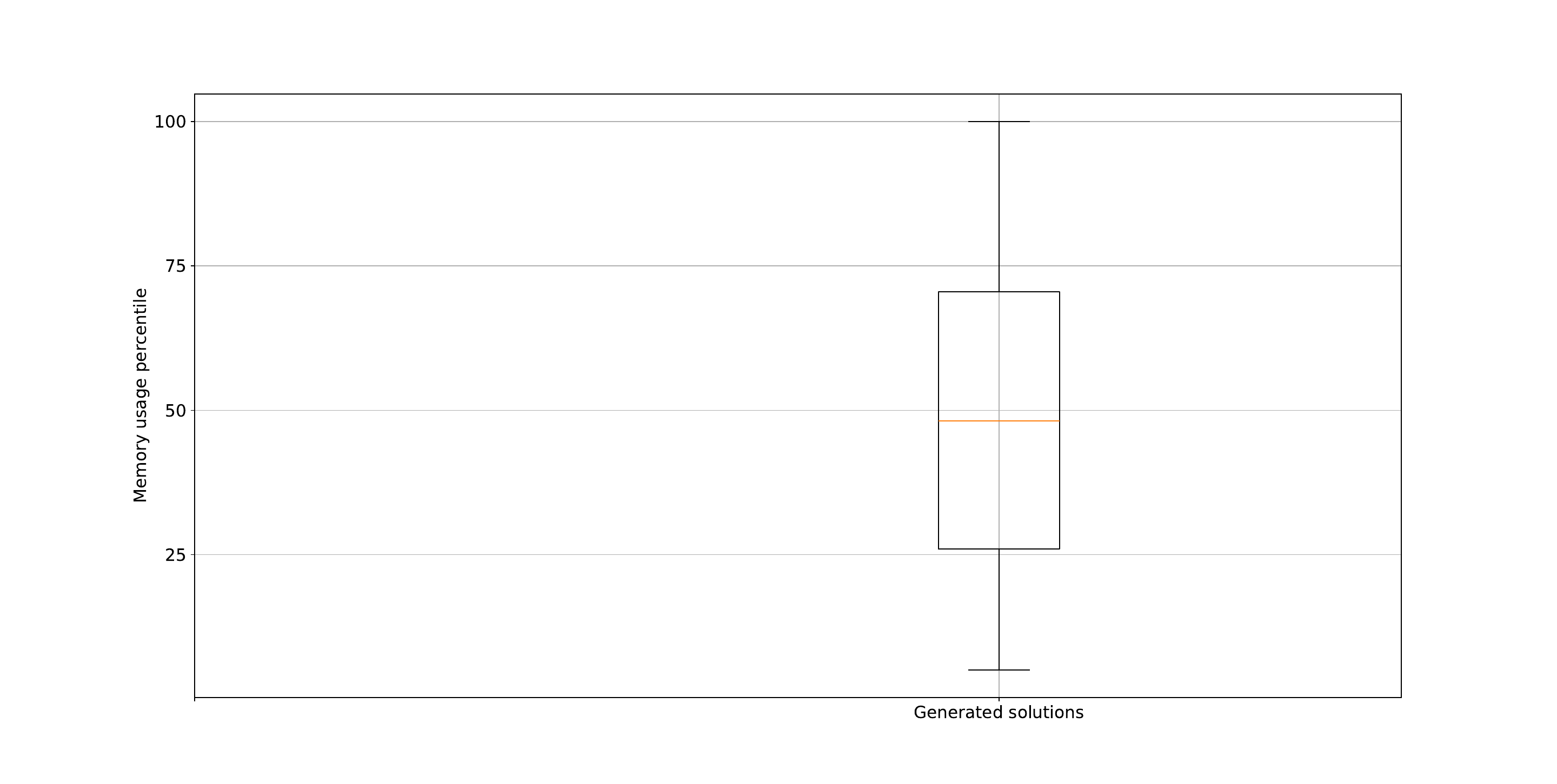}
    \vspace{-15pt}
    \caption{Boxplot of memory usage rank of the generated solutions on LeetCode}
    \label{fig:bxpl-memory-percentile}
\end{figure}

This analysis is based on the complete data set comprising 2,086 coding problems for which a valid solution could be generated. As illustrated in Table \ref{tab:RessourcesTotal}, the median and mean values are below the 50th percentile. This leads to the conclusion that the median ranking of the solutions provided by users is higher and that they consume fewer resources than the generated solutions. It should be noted that this statement does not apply to the problems pertaining to the difficulty level designated as \textit{Hard}. This is evident from Table \ref{tab:RessourcesHard}, which shows that both the median and the mean value are just above the 50th percentile. This indicates that the other difficulty levels exert a downward influence on the overall mean. In the boxplot in Figure \ref{fig:bxpl-memory-percentile}, it is notable that the IQR and the whiskers cover almost the entire range from 0 to 100, with the boxplot positioned almost in the centre.

The lower mean value of less than 50 indicates that the null hypothesis cannot be rejected. As the data set represents a non-normally distributed population, the Wilcoxon test was performed. With a p-value of $5.85 \times 10^{-02}$, this is not below the significance level of $1.25 \times 10^{-02}$, which means that the null hypothesis cannot be rejected.

\textbf{Result for H$_{1}^{3}$:} GenAI produces code that utilises equal or more resources than developers on LeetCode.

\subsection{H$_{1}^{4}$ GenAI produces code that takes less time to run than developers on LeetCode.}\label{sec:H14}
The final hypothesis pertaining to the initial research question concerns the time behaviour of the submitted solutions on the LeetCode platform. The evaluation is based on the \textit{runtime rank} on LeetCode, which enables a comparison with other users. All pertinent information is presented in Table \ref{tab:Runtime}, including the mean and median runtime rank.

\begin{table}[h]
    \centering
    \begin{tabular}{lllllll}
        \toprule
        \textbf{Samples} & \multicolumn{1}{c}{\textbf{Problems}} & \multicolumn{1}{c}{\textbf{Mean}} & \multicolumn{1}{c}{\textbf{Median}}\\
    \midrule
    GPT-4o solutions & \multicolumn{1}{r}{2,086} & \multicolumn{1}{r}{55.73} & \multicolumn{1}{r}{57.18}\\
    \bottomrule
    \end{tabular}
    \caption{Runtime rank of generated soltuions on LeetCode}
    \label{tab:Runtime}
    \vspace{2mm}
    \footnotesize{Mean and median are presented in \%}
\end{table}

\begin{figure}[ht]
    \centering
    \includegraphics[width=\columnwidth]{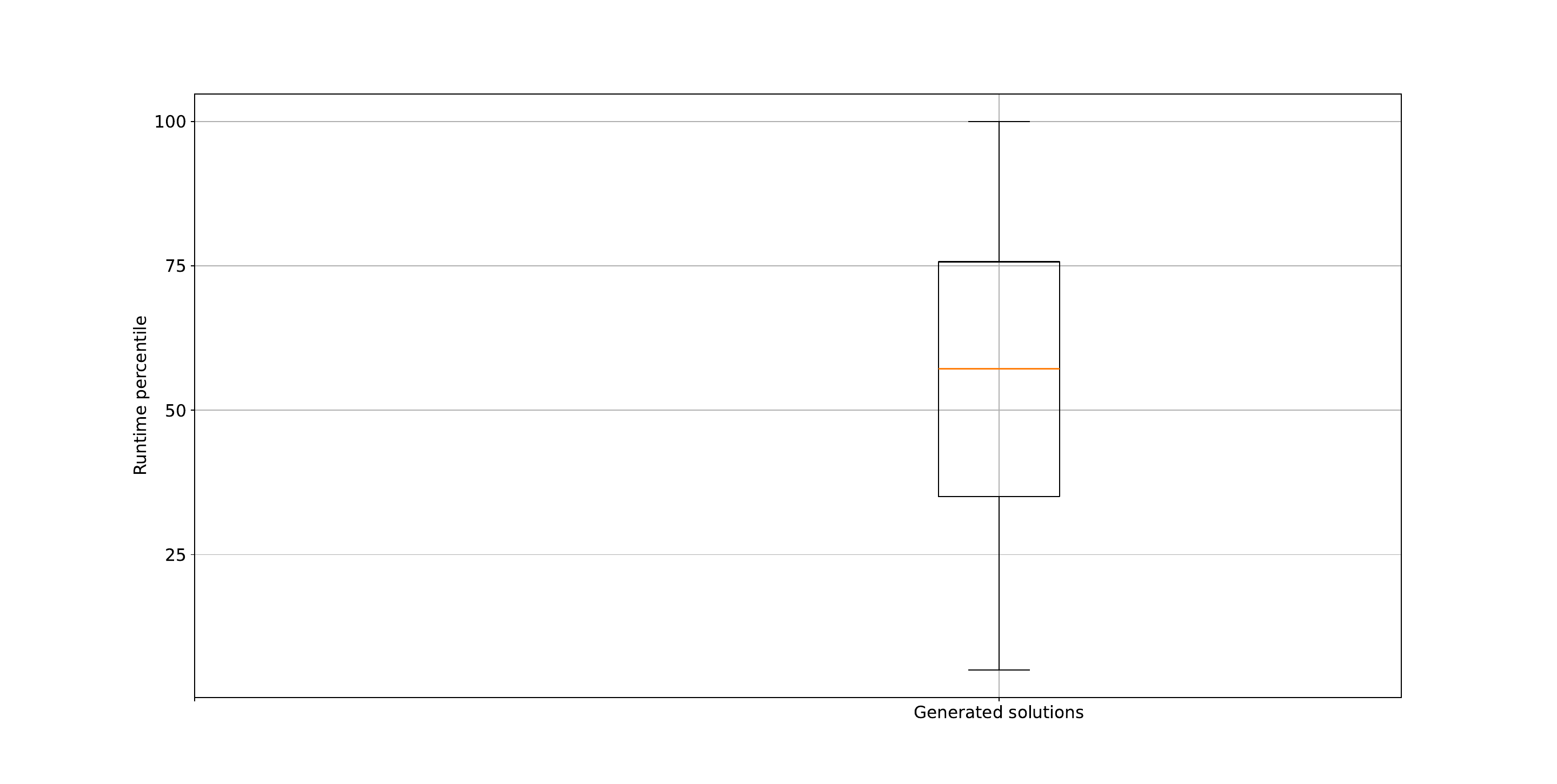}
    \vspace{-15pt}
    \caption{Boxplot of runtime rank of the generated solutions on LeetCode}
    \label{fig:bxpl-runtime-percentile}
\end{figure}

The mean and median values are both above the 50th percentile, at 55.73 and 57.18, respectively, for the 2,086 programming problems. This suggests that the mean runtime of the generated solutions is less than that of the users on LeetCode. As illustrated in Figure \ref{fig:bxpl-runtime-percentile}, the IQR is not symmetrical around the value 50, but is slightly higher. 

The data corroborate the formation of the alternative hypothesis. In the absence of a normal distribution of the data set, the Wilcoxon test was conducted. The p-value of $1.17 \times 10^{-22}$ is below the adjusted significance level of $1.25 \times 10^{-02}$, indicating that the null hypothesis can be rejected and the alternative hypothesis accepted. Nevertheless, the d-value of 0.44 reported by Cohen indicates a small effect.

\textbf{Result for H$_{1}^{4}$:} Code solutions generated by GPT-4o take a significantly lesser time to run in comparison to those developed by human coders on LeetCode. Consequently, the time behaviour of the generated code by GPT-4o is superior when confronted with LeetCode problems.

\subsection{Summary}
Table \ref{tab:RQ1Results} provides an overview of the hypotheses pertaining to the initial research question. In each case, the alternative hypothesis is indicated, and whether the hypothesis was accepted. Additionally, the table lists the p-values and test statistics determined by Python. If the hypothesis was accepted, the effect size is also indicated. It can thus be concluded that three of the four null hypotheses can be rejected, and the respective alternative hypothesis can be accepted. Specifically, hypotheses H$_{1}^{1}$ and H$_{1}^{2}$ regarding code quality and code undestanderbility were accepted with medium and small effect sizes, respectively. For the hypotheses regarding performance efficiency, only H$_{1}^{4}$ regarding time behaviour could be accepted; hypothesis H$_{1}^{3}$ on resource utilisation could not be accepted.

\begin{table}[h]
    \centering
    \begin{tabularx}{\linewidth}{p{4.8cm} XXllll} 
    \toprule
    \textbf{Alternative Hypothesis} & \multicolumn{1}{c}{\textbf{Test Statistics}} & \multicolumn{1}{c}{\textbf{p-Value}}  & \multicolumn{1}{c}{\textbf{d$_{Cohen}$}} & \multicolumn{1}{c}{\textbf{Accepted}}\\

    \midrule
    & \\[\dimexpr-\normalbaselineskip+2pt]
    \textbf{H$_{1}^{1}$} GenAI produces better code quality than developers on LeetCode. & \multicolumn{1}{r}{1,391,207.5} & \multicolumn{1}{r}{$2.13\times 10^{-89}$} & \multicolumn{1}{r}{0.65} &  \multicolumn{1}{c}{Yes} \\\hline
    & \\[\dimexpr-\normalbaselineskip+2pt]
    \textbf{H$_{1}^{2}$} GenAI produces better code understandability than developers on LeetCode. & \multicolumn{1}{r}{1,994,182.0} & \multicolumn{1}{r}{$4.01 \times 10^{-06}$} & \multicolumn{1}{r}{0.14} & \multicolumn{1}{c}{Yes}\\\hline
    & \\[\dimexpr-\normalbaselineskip+2pt]
    \textbf{H$_{1}^{3}$} GenAI produces code that utilises less resources than developers on LeetCode. & \multicolumn{1}{r}{1,034,307.5} & \multicolumn{1}{r}{$5.85 \times 10^{-02}$} & \multicolumn{1}{r}{-} & \multicolumn{1}{c}{No}\\\hline
    & \\[\dimexpr-\normalbaselineskip+2pt]
    \textbf{H$_{1}^{4}$} GenAI produces code that takes less time to run than developers on LeetCode. & \multicolumn{1}{r}{817,996.0} & \multicolumn{1}{r}{$1.17 \times 10^{-22}$} & \multicolumn{1}{r}{0.44} & \multicolumn{1}{c}{Yes}\\
    \bottomrule
    \end{tabularx}
    \caption{Overview of hypothesis results from RQ1}
    \label{tab:RQ1Results}
\end{table}

\section{Discussion}\label{sec:discDM}
The null hypothesis was rejected for three of the four hypotheses, namely with regard to code quality, code understandability and time behaviour when executing the code. However, contrary to expectations, the null hypothesis regarding resource utilisation could not be rejected. The results pertaining to code quality and understandability will now be discussed. As the time behaviour and resource utilisation characteristics fall within the domain of performance efficiency, they will be addressed in a subsequent step.

\subsection{Code Quality and Understandability}
In terms of code quality and code understandability, there is a possibility that the assumption that GPT-4o exhibits a higher standard of software quality than an average developer due to the high proportion of training data~\cite{villalobos2024rundatalimitsllm}. It was previously hypothesised that GPT-4o has the capacity to develop a more optimised solution for coding problems that surpasses the capabilities of developers on LeetCode. However, this raises the question of whether the data set is sufficiently meaningful and whether further factors need to be considered that could influence the results.

It is, however, conceivable that the quality of the software provided by the user on LeetCode may be perceived as inferior by developers with less expertise. This can be attributed to the fact that the platform serves not only to enhance programming expertise, but is also utilised by programmers preparing for job interviews. This observation leads to the assertion that the platform is becoming increasingly utilised by developers with diminished practical experience, who typically do not produce software of the same quality as experienced developers~\cite{8166711}, or the GPT-4o model. In light of the challenges associated with assessing the proficiency of developers on LeetCode, a suboptimal hypothesis is put forth: as the intricacy of the programming challenges rises, there is a heightened likelihood that these can be addressed in a valid manner by experienced developers. Conversely, those with less experience are more prone to encounter difficulties when attempting to solve more challenging programming problems. To test this assumption, it is proposed that more experienced developers generate a smaller number of code smells than less experienced developers. Therefore, a potential correlation between the number of code smells per line of code and the level of difficulty is examined. As the data is ordinal in nature (with categories for difficulty: \textit{easy, medium, difficult}), the Spearman rank test was carried out in Python. This yielded a significant result, albeit with a weak correlation (p-value = $1.48 \times 10^{-31}$, $\rho = -0.25$). This indicates that the number of code smells per line of code generally decreases as the difficulty of the problems increases. This can be interpreted in two different ways. It is plausible that the data may reflect a disparity in proficiency levels among developers, indicating the presence of a more extensive spectrum of expertise. This assumption appears to be reasonable, given that less experienced developers may lack the opportunity to address more complex problems, and even experienced developers seek a challenge from LeetCode. Conversely, the correlation is only weakly pronounced, and thus does not correspond to a perfect monotonic distribution that would underpin the assumption. Furthermore, the assumption that experienced developers produce fewer code smells seems suboptimal. In any case, the varying degrees of difficulty among the coding problems indicate that the platform will also appeal to experienced developers. Unfortunately, there is no way to examine the experience of the developers and determine whether we consequently demonstrate a lower level of software quality than in other contexts due to the users of the LeetCode platform. Further research is required in this context.

It is also important to consider that the age of the coding problems can have an impact on software quality. As previously stated, GPT-4o demonstrates a constrained capacity for generalization (also shown by~\cite{Liu}), which may also impact the quality of the software. Consequently, an investigation was conducted to ascertain whether the two metrics correlate with the age of the problem on LeetCode, specifically whether the quality of newer problems deteriorates or not. The Spearman rank correlation tests was performed for both metrics, yielding a statistically significant result, albeit with a very minimal correlation and thus no evidence of a monotone relationship in the distribution of the data (code quality: p-value = $8.97 \times 10^{-4}$, $\rho = 0.073$; code understandability: p-value = $7.24 \times 10^{-4}$, $ \rho = -0.074$). It can thus be inferred that the age of the problems has no impact on the quality of the generated code or its complexity. This observation indicates that the valid solved \textit{after-problems} exhibit a comparable quality to the \textit{before-problems}. On initial examination, this appears to be at odds with the finding that there was no significant lower values in the cognitive complexity score per LOC for the generated solutions for the \textit{after-problems} in comparison to the solutions produced by the users. Nevertheless, the correlation test merely serves to reinforce this observation and offers further insights. It should be noted, however, that the sample of \textit{after-problems} is relatively small, comprising only 107 cases. Moreover, the number of user solutions in this period is restricted. In this study, the deadline for the collection of posts was reached in mid-March 2024, a mere 4.5 months after the cutoff for GPT-4o's training data. The number of posts is limited, which may indicate a concentration of more experienced developers seeking new challenges and therefore providing less complex code. These are, however, speculative assumptions. Nevertheless, the hypothesis test and the correlation test serve to illustrate the multitude of factors that exert an influence on the quality of software. Notably, GPT-4o produces a fairly consistent quality of valid coded problems over time.

\subsection{Performance Efficiency}
With regard to performance efficiency, namely the utilisation of resources and the time behaviour, only the latter exhibited a significant result. However, the observed practical effect was, as indicated by Cohen's d, only small. Both metrics were in the median around the 50th percentile, with the memory rank slightly below (48.16\%) and the runtime rank slightly above (57.18\%). This suggests that GPT-4o's value for both metrics is average and that there is only a practical, small difference for the runtime rank.

One potential explanation for this discrepancy may be that, in contrast to the extensive literature on code quality and complexity on the Internet, there may be a paucity of documentation on the optimisation of memory usage. To illustrate, the \textit{KISS}\footnote{KISS, \url{https://medium.com/@curiousraj/the-principles-of-clean-code-dry-kiss-and-yagni-f973aa95fc4d}} (Keep It Simple, Stupid) principle demonstrates the significance of clean code. In light of this principle, it is recommended that the code be designed in an easily understandable and maintainable form wherever feasible. In such cases, the optimisation of the utilisation of resources is often not the primary objective, provided that it facilitates the comprehension of the code. However, the presented assumptions do not explain the significantly superior runtime values, while the memory usage is below average. Moreover, there are entire research areas that specialise in system performance, with entire lectures dedicated to the subject. An illustrative example can be found in the field of theoretical computer science, where the time and space hierarchy classification plays a pivotal role.

An additional potential explanation for the discrepancy between runtime and memory usage efficiency is the presence of a greater number of unused objects, such as lists, dictionaries, or other collections, within the GPT-4o code in comparison to human-written code. This is a rule established by SonarQube, which is presented in the form of a code smell. Nevertheless, an analysis of the code smell is not feasible given that only the code of the generated solutions is accessible, not that of the submitted user solutions. In this context, the published code in the posts represents only a limited portion of the total code of the user solutions. Furthermore, the initial hypothesis posits that the code generated exhibits a markedly diminished prevalence of code smells per line of code in comparison to code written by humans. This observation renders the assumption that the unused objects are an essential feature implausible. This line of reasoning can also be applied to the assumption that the generated code is simply longer or more complex. The latter assumption is inconsistent with the second hypothesis (cognitive complexity per LOC), while the former is challenging to quantify, as previously discussed in the context of code smells. A limited investigation indicated that the average length of the code for both sides was relatively balanced, at 16.8 lines for the generated solutions and 15.8 lines for the solutions from thr posts. This makes the argument of longer code implausible.

An additional rationale may be that the solutions on LeetCode are of an exemplary standard with regard to both metrics. This is due to the fact that LeetCode displays the corresponding metrics to each user, which provides an incentive for users to develop solutions that are more efficient than those of their peers. The objective is therefore to develop solutions that consistently outperform those of other users with regard to both metrics. In numerous posts, data was presented indicating the number of users who had been surpassed, which serves to corroborate this assumption. It may therefore be assumed that GPT-4o is unable to provide an even more optimal solution regarding the memory usage than the already highly optimised solutions provided by users. Nevertheless, this does not elucidate why the generated code exhibits a markedly reduced runtime, yet only a higher average memory usage compared to the users from LeetCode. 

A final potential explanation for this discrepancy is that the code in the training data prioritised runtime over memory usage. In the context of algorithmic problems, a shorter runtime is often the primary objective for users, while memory usage is frequently regarded as a secondary concern. Further investigation into the code domain and the influence of runtime and memory usage is necessary to elucidate the underlying causes of these observed differences. It is noteworthy that the methodology employed by Siddiq et al.~\cite{Siddiq2022}, which involved an examination of the training data of an open-source GenAI with a focus on code smells, could be replicated with a similar approach to assess performance efficiency.

In regard to the markedly superior runtime of the generated code in comparison to that of the developers on LeetCode, this study corroborates the findings of Coignion et al.~\cite{Coignion}. However, it should be noted that Coignion et al. only collected data on runtime, and not on memory usage. Consequently, a comparison on the latter is not possible.

\subsection{Implication}
It can be posited that the utilisation of GPT-4o does not impede the software developer in terms of code quality and complexity, nor does it affect runtime behaviour when generating code in Python. Consequently, it \textit{may} not negatively impact the software product when employed on a limited scale. Nevertheless, there remain numerous unanswered questions that have the potential to affect software quality and cast the data set of this study in a different light. Therefore, further research is necessary in this regard, and the results of this study should be contextualised within the framework of this research.

\section{Conclusion and Outlook}\label{chap:conclusionAndOutlook}
This study illustrates the transformative potential of GenAI in the context of computer science. In order to elucidate this hitherto obscure research area, the fundamental principles of the software quality of generated code and a comparison with the quality of code written by humans were initially presented. As part of this endeavour, a comprehensive web scraping and data mining process was undertaken to obtain a problem set comprising 2,321 coding problems from LeetCode. In the course of this process, 2,086 valid solutions were generated by OpenAI`s GPT-4o, while 57,238 valid solutions were collected by users of the LeetCode platform. A total of 958,574 lines of code in Python were analysed to ascertain whether the software quality of the GPT-4o model is superior to that of an average developer on LeetCode. The evaluation of software quality is based on four main criteria: code quality, code understandability, time behaviour and resource utilisation. Static code analysis was conducted using SonarQube to assess the quality of the software in terms of code smells and cognitive complexity. The LeetCode platform provided data regarding the memory usage and the runtime of the generated code. 
The objective of this study was to ascertain whether ChatGPT offers beneficial assistance to software engineerer in addressing software-related tasks or, conversely, impedes their capabilities. 

The investigation of the programming problems from LeetCode has revealed that GPT-4o exhibits deficiencies in its capacity to generalise. This results in a diminished performance of the model when attempting to solve tasks that are not included in the training data. The proportion of instances in which the problem was solved correctly within five attempts was 51.94\%. However, it is noteworthy that the general solution rate of all publicly and freely available problems on LeetCode was 89.88\%. A comparison of the valid solution with that of human-written code revealed that GenAI generally produces code of a higher quality than humans. However, this assertion is limited to the assessment of code quality in terms of the number of code smells per LOC (medium effect), the code understandability through the cognitive complexity score per LOC (small effect), and the time behaviour through the runtime rank (small effect). Contrary to expectations, only the resource utilisation by the memory usage rank shows a different picture, with the LeetCode developer showing superior quality.

The study introduces a novel methodology for the comparison of handwritten code with generated code on LeetCode. This provides a framework for other studies to replicate and expand upon this results, thus facilitating further research in this field. In accordance, the code and data set have been made accessible via the Zenodo\footnote{Zenodo repository with scripts and data, \url{https://doi.org/10.5281/zenodo.13881451}} repository. The concept of software quality is complex and incorporates a range of elements that extend beyond the four metrics presented in this context. It is thus imperative that future research endeavors to gain a more comprehensive understanding of the diverse range of quality characteristics inherent to generated code. Furthermore, it is vital to examine user experiences on LeetCode in order to gain a comprehensive understanding of the quality of the software produced by them. Accordingly, the data from this study can be organised and the quality generated can be classified in accordance with the user experiences. Moreover, a comparison of the generated code with that produced by users in different programming languages and using different LLMs would be a valuable addition to the findings of this study. A foundation has already been established in this study, and a substantial corpus of data has been assembled for use in future studies. 

\backmatter








\section*{Declarations}


\subsection*{Funding}
This research received no external funding. This article was funded by the Open Access Publication Fund of the Federal Institute for Vocational Education and Training (BIBB), Bonn.

\subsection*{Conflict of interest/Competing interests}
The authors declare no conflict of interest.

\subsection*{Ethics approval and consent to participate}
Not applicable.

\subsection*{Consent for publication}
Not applicable.

\subsection*{Data availability}

Zenodo repository with scripts and data, \url{https://doi.org/10.5281/zenodo.13881451}.

\subsection*{Materials availability}
Not applicable.

\subsection*{Code availability}

Zenodo repository with scripts and data, \url{https://doi.org/10.5281/zenodo.13881451}.

\subsection*{Author contribution}








\begin{appendices}

\section{Problem Requirements}\label{chap:app:pr}
Upon completion the retrieval of the coding problems, the foundation of this problem set was established. The selection was limited to problems related to the Python programming language, given its widespread popularity (according to a Stack Overflow survey\footnote{Stack Overflow survey, \url{https://survey.stackoverflow.co/2024/technology}}). The language is distinguished by its intuitive syntax, dynamic typing, and interpreted execution~\cite{PythonProgramming}. The popularity of Python, which is also embraced by beginners due to its intuitive nature, has led to its pervasive use as a programming language. This may have attracted a considerable number of individuals with an interest in providing solutions on the LeetCode platform. In the light of the aforementioned considerations, Python was selected as the programming language for analysis. Subsequently, 81 of the total 2,992 individual problems were excluded from the data set as they could not be solved with the Python programming language, only with SQL (category \textit{Databases}) and Bash (category \textit{Shell}). As previously stated in step one, there are also coding problems that can only be viewed and solved with a premium plan at LeetCode. Consequently, a further 590 premium problems had to be excluded, leaving 2,321 individual programming problems for the evaluation of the GenAI. In some studies, the coding problems were sorted out prior to the date of the last training data set of the AI in order to prevent contamination of the data~\cite{jacovi2023stopuploadingtestdata}. This approach was not adopted, as a comparison of the generated solutions to the problems included in the training data with those created by humans was also of interest, as this will yield insight into the quality of the AI's output. However, the analysis was also been divided into problems that are deployed prior to and subsequent to the cutoff date of the training data, in order to ascertain the extent of the AI's generalisation abilities.

\section{Generate Solutions}\label{chap:app:gs}

\textbf{1. Request problem information from LeetCode.} The first step entails the retrieval of supplementary data from LeetCode for the purpose of formulating a prompt: the problem descriptions and the Python framework in which the solution was to be implemented. 
A GraphQL query was executed on the resource \textit{questionContent} using the parameter \textit{titleSlug} (the title of the problem, with a hyphen used instead of spaces) in order to retrieve the problem description. The \textit{questionEditorData} resource was queried with the \textit{titleSlug} parameter in order to obtain the code structure.

\textbf{2. Assemble the input for the API call.} In the second step, a parallel iteration is conducted across all problem definitions, i.e. the description and the code framework, with the objective of assembling the prompt. The objective is not to consistently identify the optimal solution through the utilisation of sophisticated prompt designs; instead, it is to construct a query that closely approximates the actual scenario. An exemplar of the prompt can be observed in Figure \ref{fig:firstPrompt}. The prompts were developed in accordance with the standards established by the model's creators\footnote{OpenAI Prompt Engineering, \url{https://platform.openai.com/docs/guides/prompt-engineering}}, as well as through a review of analogous studies for comparative purposes~\cite{tian2023chatgptultimateprogrammingassistant, Liu, LiuSQ}.

The structure of the prompt was consistent across all problems:
\begin{enumerate}
    \item The precise and detailed command\footnote{Clear Command Tactic, \url{https://platform.openai.com/docs/guides/prompt-engineering/strategy-write-clear-instructions}} for the GenAI containing the specifications for the generation of the solution. It was determined that the sole objective of the generative GenAI should be the generation of Python code, as any accompanying description would serve no practical purpose.
    \item The problem description, which also contained the examples and the parameter restrictions, was enclosed in three quotation marks\footnote{Delimiters Tactic, \url{https://platform.openai.com/docs/guides/prompt-engineering/tactic-use-delimiters-to-clearly-indicate-distinct-parts-of-the-input}} to distinguish it from the command.
    \item The second clear command, which instructed the AI to use the code framework defined in the next point. The command was separated from the first command to split\footnote{Split Tasks Tactic, \url{https://platform.openai.com/docs/guides/prompt-engineering/strategy-split-complex-tasks-into-simpler-subtasks}} the problem and create a clearer structure for the AI.
    \item The code framework served as the framework for the AI to generate solutions. This code block was initiated with three backticks and the programming language to indicate the commencement of a code sequence and terminated with three backticks to demarcate the end of the code block\footnote{Code Block Tactic, \url{https://platform.openai.com/docs/guides/prompt-engineering/tactic-use-code-execution-to-perform-more-accurate-calculations-or-call-external-apis}}. This delineation allows the AI to discern the code within this block and its associated programming language.
\end{enumerate}

\textbf{3. Generate the code via OpenAI API.} In the third step of the process, interaction with the LLM was initiated. The \textit{gpt-4o-2024-05-13}\footnote{Model GPT-4o, \url{https://platform.openai.com/docs/models/gpt-4o}} model, representing the latest version of OpenAI at the time of the study, was employed. The model has a knowledge cutoff date of October 2023. An API for Python is provided by OpenAI, which facilitates the automation process. Furthermore, the provided API enabled the specification of parameters\footnote{API Parameters for Chat, \url{https://platform.openai.com/docs/api-reference/chat/create}} related to the model, in addition to the model and prompt. In the evaluation of the LLM, the temperature of the model is frequently subjected to experimentation, e.g.~\cite{Doderlein}. The temperature serves as a parameter that delineates the deterministic (lower temperature value) or random (higher temperature value) characteristics of the output. However, in order to ensure the greatest possible degree of real-world applicability, this study retains the default value of 1, which is in the middle of the possible spectrum (0 and 2) provided by OpenAI for their model\footnote{Temperature OpenAI API, \url{https://platform.openai.com/docs/api-reference/chat/create}}. Moreover, the number of outputs to be generated by GPT can be specified. A value of \texttt{n = 1} was selected; further details on this can be found in the last step \textit{\# 6}. It is possible to set a limit for the number of tokens for input and output, but this may result in the generation of an incomplete answer and the omission of parts of the code. Accordingly, the maximum default value was also selected for this parameter, which is 4,096 tokens in total (approximately 100 tokens correspond to 75 words). The defined limit encompasses the total number of input and output tokens. Nevertheless, the value of the tokens is relatively high and has never been reached within the scope of this study. To exemplify this, the \textit{Two Sum} problem necessitated the input of 306 tokens and the output of 145 tokens. Consequently, a total of 451 tokens were used, which represents a slight excess of 10\% of the limit. In the second step assembling the prompt, a restriction was defined that prescribed the generation of code without explanations. This resulted in a reduction in output and, consequently, a reduction in token usage. Finally, by specifying the model, prompt, number of outputs, and maximum number of tokens, it was possible to generate a solution to the coding problem. In Step 2, the LLM was instructed to implement the code within the three backticks in the code framework. A regular expression was used to facilitate the extraction of the generated code, ensuring its consistent transfer to the subsequent step.

\textbf{4. Submit the code on LeetCode.} The generated code had to undergo a validation process prior to analysis. As part of the sampling process, the code was submitted to LeetCode for execution and testing using the test cases provided by LeetCode. Due to the difficulties encountered in executing the request to submit the solution using Python, an alternative approach was adopted, utilising the Selenium\footnote{Selenium, \url{https://www.selenium.dev/}} framework. Selenium is a framework designed for the automated testing of web applications. Furthermore, it is particularly well-suited to the domain of web scraping, as it facilitates the automation of browser instructions~\cite{chapagain2019hands}. In the Python script, the importation of the \textit{selenium}\footnote{Selenium library, \url{https://pypi.org/project/selenium/}} library was conducted, thereby enabling the full sequence of browser actions to be programmed. The initial step involved accessing the LeetCode page associated with the specific problem and establishing a valid session by setting cookies. This circumvented the conventional sign-up process, which would have entailed solving captchas automatically, a challenging task. Subsequently, the \textit{Python3} programming language could be selected in the LeetCode IDE, the code generated by the LLM could be inserted, and the submission could be initiated.

\textbf{5. Get information on submitted code.}
Due to the time required for the execution of the code and the evaluation of the tests associated with the submitted solution on the LeetCode platform, the feedback on the submitted solutions was obtained from LeetCode in two queries during the fifth step of the sampling process. 
\begin{enumerate}
    \item The five latest solutions were requested via a GraphQL query on the \textit{submissionList} resource, utilising the title of the problem as the \textit{questionSlug} parameter and the language Python as the \textit{lang} parameter. The final submission was identified from the list and the corresponding ID was established. 
    \item A further query on the \textit{submissionDetails} resource, with the ID parameter, provided all the requisite information about the submission of the generated solution. In addition to the information about the problem, the status of the submission (accepted or not accepted), the total number of test cases for the problem and the number of successfully executed test cases, the error information in the event of errors, and other metrics were stored. This information was then passed on to the final step of the sampling process.
\end{enumerate}

\textbf{6. Evaluate submission.}\label{Step6}
In the event that the status was designated as ``Accepted'', the solution was stored and the generation of the solution for the subsequent problem commenced. In the event that this is not the case and an error occured during the execution of the code or other issues arise, a multi-round fixing approach is employed, similar to~\cite{Liu}. This process emulates a dialogue with the GenAI, as a human user would conduct in the event of the AI incorporating errors into the generated code. In such a scenario, a new solution had to be generated through a new prompt, with further information about the error included. In contrast to the capabilities of ChatGPT, the API does not permit the submission of further queries within the same chat session. Moreover, the context is lost after a query is submitted. Accordingly, the prompt had to be substantially modified in order to generate a new solution. An illustrative example of a prompt to rectify an error in the code is presented in Figure \ref{fig:errorPrompt} and has the following structure:
\begin{enumerate}
    \item The revised command provided GPT with the specific details of the error in the code and the objective of rectifying it.
    \item The initial problem definition was reiterated, as the context was no longer accessible.
    \item The second command was then issued, announcing the error.
    \item The error message was accompanied by the relevant information from LeetCode.
    \item The third command was provided, announcing the faulty code.
    \item The erroneous code generated by the LLM that had to be amended.
\end{enumerate}

In formulating the revised prompt, due consideration was again given to ensuring compliance with the established standards for writing a prompt. Subsequently, Steps 3 to 6 were repeated with the new prompt. 
A maximum of five attempts was permitted to achieve a valid solution to the coding problem. This number has been demonstrated to be sufficient in previous studies~\cite{Bucaioni, dong2024selfcollaborationcodegenerationchatgpt}. 


\section{User Solutions}\label{chap:appendix:us}

\textbf{1. Query list of all community posts.}
Once a solution to the problem has been successfully developed and all test cases have been completed successfully on LeetCode, users are given the option of making the solution publicly available. In this regard, users are afforded the option of composing a post in Markdown\footnote{Markdown, \url{https://www.markdownguide.org/}} according to their own preferences. Additionally, it is possible to assign a title and specify tags. The initial tag is automatically assigned the name of the programming language utilised to resolve the coding problem. Subsequently, other users are afforded the opportunity to view and evaluate these posts by casting upvotes. Moreover, other users are afforded the opportunity to leave a comment beneath the post. A list of posts is provided under each respective problem. Accordingly, the initial stage of the sampling process entailed the retrieval of all posts presenting handwritten solutions, without imposing any preliminary constraints pertaining to the specific programming language or the nature of the content. However, to guarantee the quality of the contributions and, in particular, to prevent contributions that consist solely of copied and pasted code, a minimum of two upvotes was required. A single upvote may be provided by the author of the post; however, a minimum of one additional upvote is required from another user. A GraphQL request was made to the \textit{communitySolutions} resource, with the \textit{questionSlug} parameter passed in order to create lists of posts for each problem. These lists were paginated, with 50 posts displayed at a time. As part of the data collection process, general information about the posts was recorded, including the title, solution tags, post ID, number of upvotes, creation date, author's username, and other metadata. However, the content of the posts was not yet available for collection. This approach resulted in a comprehensive list of 278,397 posts, which were subsequently examined in the next step.

\textbf{2. Query each community post.}
By iterating over the list of posts from the previous step, the content of the post was requested using a query to \textit{communitySolution} with the ID of the post as the \textit{topicId} parameter. At this point, the content was subjected to its first round of validation. The code in the original post is indicated in Markdown format by enclosing three backticks. In order to extract the code from the Markdown in the subsequent step, it was necessary for users to ensure that the Markdown was correctly formatted. However, it was noted during the web scraping process that some posts did not comply with the required formatting standards, specifically the use of three backticks. Nevertheless, it was not feasible to ascertain whether the issue originated from LeetCode or from the users. In some cases, it was observed that the number of consecutive backticks did not correspond to an even number. Indeed, the \texttt{\# backtick blocks modulo 2 = 0} should per post evaluate to true, given that a user is permitted to integrate several code solutions into a single post. It should be noted, however, that the LeetCode web interface correctly displayed the code in the post, indicating that the render engine or the LeetCode editor itself is capable of recognising and automatically correcting faulty Markdown code. In such a scenario, the pertinent post could be retrieved through the utilisation of Selenium. By entering the URL \textit{https://leetcode.com/problems/\{questionSlug\}/solutions/\{postID\}/\{postTitleSlug\}}, it was possible to access the respective post directly via the browser. The markdown was obtained by searching for the class ``FN9Jv WRmCx'' in the HTML page. Consequently, the raw content could be obtained for each post ID either via a GraphQL query or by Selenium, and subsequently transferred to the subsequent step.

\textbf{3. Extract code snippets from post.}
The third stage of the process involved the extraction of all code blocks from the Markdown document and their subsequent assignment to a specific programming language. The extraction of the code was initiated only when it exceeded a length of two lines. This is due to the fact that in Python, a single line is required for a class definition, while an additional line is necessary for the definition of a function and at least one line for the implementation of a solution. In certain instances, the use of backticks was employed to elucidate the solution, for example, by providing variables within three backticks. However, this was inconsequential with respect to the ultimate code block with the full solution, and thus had not to be considered. Accordingly, the code was employed only from a length of three lines onwards. In order to ascertain the programming language of the code, Markdown allows the user to specify the relevant programming language after the initial backticks, thus facilitating the clear assignment of the code block to a specific language. In the absence of a precise definition, an examination of the tags was required. In some cases, the specific programming language employed was not specified, or multiple languages were referenced. In such instances, the title was subjected to a search for a distinctive language. In the absence of a discernible programming language, the Python library Pygments\footnote{Pygments, \url{https://pygments.org/}} was employed as a means of identifying the language used for implementation. Pygments is typically employed in the context of syntax highlighting; nonetheless, its functionality can also be applied to the recognition of a programming language from a string containing code. However, the results of the manual testing of the lexer did not meet the desired outcome. As a result, a unique lexical analyser was created. In the case of certain programming languages from LeetCode, the process involved defining keywords and patterns and then assigning a weight to each. Subsequently, the presence of these keywords and patterns of each programming language was determined for the extracted code, and a score was calculated based on this, taking the weighting into account. The programming language with the highest score was then selected. While this process did not yield perfect precision, this was not a significant issue, as an incorrect assignment could be validated in the subsequent step. However, the recall could not be determined, but this was not considered problematic due to the large number of samples.

\textbf{4. Validate user solution.}
In the final stage of the process, the non-valid code snippets were removed, and the Python code solutions were consolidated into a single solution per post and problem. In order to achieve this, a local evaluation of the code was conducted. Subsequently, in some cases, multiple code snippets with an identified programming language were present within a single post. In this manner, the code was frequently assembled in a step-by-step manner, with the final solution typically presented as the concluding element. Consequently, the code snippets were evaluated in a sequential manner, from the latter to the former, with the objective of identifying any valid code. As the objective was to examine solely Python code snippets, a comprehensive analysis of the code was conducted in cases where Python had been previously assigned to the snippet.

It should be noted that, due to the internal handling of common imports on LeetCode, users are not required to specify the individual imports of the libraries used. However, the imports were required for validation purposes in order to execute and analyse the code, as otherwise error messages will appear. The process of adding the imports was not a trivial task and required several steps:
\begin{enumerate}
    \item The \textit{autoimport}\footnote{Module autoimport, \url{https://pypi.org/project/autoimport/}} library was employed to incorporate the most common imports into the code.
    \item In many instances, the \textit{autoimport} library proved insufficient, particularly when utilising more specialised libraries. Subsequently, a static code analysis was conducted utilising the \textit{pylint}\footnote{Module pylint, \url{https://pypi.org/project/pylint/}} library. In the event of an absence of requisite imports or an absence of definition for variables, the message ``Undefined variable \{variabele name\}'' was logged.
    \item The pertinent log entries were collated and a mapping was constructed in Python to an import. This enables Python to automatically access the mapping when the same module is required for subsequent importation. However, the mapping had to be created manually and ultimately comprised 89 different imports.
    \item The missing imports had to be correctly inserted in the code before the class definition and in the appropriate sequence with the correct spacing. This part was also automated.
\end{enumerate}

Subsequently, the code incorporating the aforementioned imports was subjected to parsing (utilising the \textit{ast} module\footnote{ast module, \url{https://docs.python.org/3/library/ast.html}}), compilation and execution. This process enabled the removal of code segments that were identified as erroneous. Moreover, this process ensured the collection of namespaces for all variables and functions defined within the code. In order for the LLM to generate solutions, it was necessary to provide the code framework from LeetCode within which these solutions were to be generated. At this stage, the code framework was reused and also parsed, compiled and executed, thus ensuring the collection of namespaces. This enabled an analysis of the code snippets in terms of their compliance with the classes and methods specified by LeetCode. This, in turn, permitted the removal of code snippets that provided only partial solutions or solutions to other problems.



\end{appendices}


\bibliography{bibliography}

\end{document}